\documentclass[prd,aps,a4paper,superscriptaddress,twocolumn,nofootinbib,floatfix]{revtex4-2}

\pdfoutput=1
\usepackage{graphicx}
\usepackage{color}
\definecolor{RED}{rgb}{1,0,0}
\usepackage{amsmath}
\usepackage{amssymb}

\allowdisplaybreaks
\usepackage[linktoc=none]{hyperref}
\hypersetup{colorlinks,			
	linkcolor=blue,
	citecolor=blue}
\begin{document}

\title{Can Oscillatory and Persistent Nonlinearities Be Bridged in Black Hole Ringdown?}

\author{Jun-Xi Shi}
\thanks{Email address: shijunxi23@mails.ucas.ac.cn}
\affiliation{International Centre for Theoretical Physics Asia-Pacific, University of Chinese Academy of Sciences, 100190 Beijing, China}


\author{Zhen-Tao He}
\affiliation{School of Physical Sciences, University of Chinese Academy of Sciences, Beijing 100049, China}

\author{Jiageng Jiao}
\thanks{Contact author: jiaojiageng@ucas.ac.cn}
\affiliation{International Centre for Theoretical Physics Asia-Pacific, University of Chinese Academy of Sciences, 100190 Beijing, China}

\author{Jing-Qi Lai}
\affiliation{School of Physical Sciences, University of Chinese Academy of Sciences, Beijing 100049, China}

\author{\protect\linebreak Caiying Shao}
\thanks{Contact author: shaocaiying@ucas.ac.cn}
\affiliation{School of Physical Sciences, University of Chinese Academy of Sciences, Beijing 100049, China}

\author{Yu Tian}
\thanks{Contact author: ytian@ucas.ac.cn}
\affiliation{School of Physical Sciences, University of Chinese Academy of Sciences, Beijing 100049, China}
\affiliation{\textit{Institute of Theoretical Physics, Chinese Academy of Sciences, Beijing 100190, China}}


\author{Hongbao Zhang}
\thanks{Contact author: hongbaozhang@bnu.edu.cn}
\affiliation{School of Physics and Astronomy, Beijing Normal University, Beijing 100875, China}
\affiliation{Key Laboratory of Multiscale Spin Physics, Ministry of Education, Beijing Normal University, Beijing 100875, China}
\date{\today}

\begin{abstract}
Quadratic quasinormal modes (QQNMs) and the Christodoulou memory effect are key nonlinear phenomena in gravitational wave physics. QQNMs characterize the near-zone nonlinear response of a perturbed black hole, whereas the memory effect is a nonlinear remnant imprinted at null infinity by outgoing radiation. This naturally raises the question of whether and in what sense the two can be bridged. We show that they are related through bridge coefficients that depend primarily on remnant black hole parameters during ringdown. Future space-based gravitational wave detectors 
can probe this relation. These results provide a new avenue for testing gravity and a fresh perspective on the nonlinear regime of general relativity.

\end{abstract}
\maketitle

\section{Introduction}
\label{sec:intro}
Gravitational waves (GWs) from the black-hole ringdown phase are crucial probes of extreme gravitational fields \cite{test:rd1,test:rd2,test:rd3,test:rd4,test:rd5,test:rd6}. In this regime, oscillatory nonlinear effects, manifested as QQNMs\kern0.3em\cite{Nakano2007,nakano2,Mitman2022,Cheung:2022}, and persistent nonlinear effects, represented by the Christodoulou GW nonlinear memory effect\kern0.3em\cite{Christ:1991,Wiseman:1991,Favata2010MemoryReview}, are the two key nonlinear signatures. Nonlinear tails are another important ringdown nonlinearity and are left for future work \cite{Cardoso2024,Ma2024,Ling2025,Kehagias2025,He:2025,Shao:2026}. Yet current ringdown observations are still largely interpreted within linear perturbation theory, even though binary black hole (BBH) mergers are intrinsically nonlinear and their spacetime response is expected to exhibit rich nonlinear structure. It is therefore natural and important to shift attention toward testing and characterizing these nonlinear effects in the ringdown phase.

QQNMs arise at second order and are driven by quadratic combinations of linear quasinormal modes (QNMs), producing modes whose frequencies are sums of the parent-mode frequencies, with amplitudes set by first-to-second-order amplitude ratios. Recent analytical and numerical studies indicate that these ratios depend on the intrinsic parameters of the remnant black hole \cite{Yuste:2024,Cheung:2024,Ma:2024,Zhu:2024,Buccio:2024,Bourg:2025,Khera:2025}. QQNMs can contribute non-negligibly to the observed ringdown signal, and there are now claims of tentative evidence for their presence in real GW data\kern0.3em\cite{Yang:2025, Wang:2026rev}.

The Christodoulou memory effect is a nonlinear phenomenon in which a burst of radiation leaves a permanent displacement between freely falling test masses, with an amplitude proportional to the square of the GW strain (or, equivalently, to the time-integrated energy flux), reflecting a second-order back-reaction of the wave’s energy–momentum on spacetime\kern0.3em\cite{Christ:1991,Wiseman:1991,Favata2010MemoryReview}. It forms one corner of the gravitational “infrared triangle”, being in one-to-one correspondence with the Bondi–Sachs asymptotic structure and BMS supertranslations at null infinity and equivalent, in the zero-frequency limit, to soft-graviton theorems\kern0.3em\cite{StromingerZhiboedov2016,Strominger2017LecturesIR}. Although an unambiguous single-event detection is still missing\kern0.3em\cite{Cheung:2024memory,PhysRevD.101.023011,Lasky:2016,Zhao:2021,Ebersold:2020}, searches in LIGO–Virgo–KAGRA data, together with sensitivity forecasts for pulsar timing arrays \cite{BabakPTA:2024} and future space-based detectors, indicate realistic prospects for a first observation in upcoming experiments\kern0.3em\cite{Incha:2024,Ghosh:2023}.

Focusing on black-hole ringdown, we consider nonlinearities that arise at second order from quadratic coupling of QNMs. 
In this framework, QQNMs describe an oscillatory nonlinear response governed by the strong-field near-zone, whereas the memory corresponds to a persistent nonlinear imprint of the outgoing radiation at null infinity. This naturally raises the question of \textit{whether these two nonlinear effects, which originate from the same sources but are tied respectively to the near-zone and to null infinity, share an intrinsic relation}.

In this work, we establish such a quantitative relation in the ringdown regime. We show that, within a given QNM coupling channel, the total accumulated displacement memory is proportional to the amplitude of the corresponding QQNM through a single multiplicative bridge coefficient. This coefficient depends mainly on the frequencies of the source QNMs and on the first-to-second-order amplitude ratio, and can therefore be predicted from the geometry of the remnant black hole. In this sense, the total ringdown memory can be expressed as a sum of bridge-coefficient–weighted QQNM amplitudes. We test this relation in the dominant coupling channel using the SXS catalog across a broad range of remnant spins, finding that the extracted bridge coefficients agree with the theoretical prediction within uncertainties. 

Using the bridge coefficient, we enable a new class of consistency tests of general relativity that jointly exploit these two nonlinear phenomena. Since QQNMs are typically more accessible than memory, they can provide informative priors for memory searches. Recent studies suggest that future space-based GW detectors could probe this QQNM–memory relation\kern0.3em\cite{Incha:2024}.
More broadly, joint measurements of oscillatory and persistent nonlinear effects open a new avenue for testing general relativity in the nonlinear strong-field regime and offer a complementary view of its nonlinear dynamics.

This paper is organized as follows: Sections \ref{sec:IIA} and \ref{sec:IIB} briefly review QQNMs and memory effect. Section \ref{sec:IIC} establishes a quantitative bridge between QQNMs and memory in the ringdown regime. In Section \ref{sec:verify}, we test the resulting bridge coefficients using numerical relativity waveforms. In Section~\ref{sec:detection}, we discuss prospects for testing the bridge relation with future space-based detectors and estimate the relative magnitude of QQNM-sourced memory compared to the parent-mode memory.
Section \ref{sec:conclusion} summarizes our main results, and discusses observational prospects and the potential role of this relation in the future.

\section{Theoretical framework}
\label{sec:theory}
Both the QQNM and the Christodoulou memory effect in the ringdown originate predominantly from quadratic interactions of linear order QNMs, suggesting an intrinsic connection between them. In this section, we begin by introducing the physical picture of QQNMs and provide a thorough discussion of the amplitude ratio between quadratic and linear order. We then describe the memory effect and its associated computational methods. Finally, we present a detailed theoretical derivation to establish the relationship between these two phenomena.

\subsection{Quadratic quasinormal modes}
\label{sec:IIA}
In the framework of black hole perturbation theory (BHPT), we can expand the spacetime metric $\tilde{g}_{\mu \nu}$ about a background $g_{\mu \nu}$ to second order in a small bookkeeping parameter $\epsilon$
\begin{equation}
    \tilde{g}_{\mu \nu} = g_{\mu \nu} + \epsilon h^{(1)}_{\mu \nu} + \epsilon^2 h^{(2)}_{\mu \nu},
\end{equation}
where $h^{(n)}_{\mu \nu}$ denotes the $n$-th order perturbation, and the corresponding expansion of the vacuum Einstein tensor yields
\begin{equation}
\begin{aligned}
    G_{\mu\nu}[\tilde{g}] = \epsilon G^{(1)}_{\mu\nu}[h^{(1)}] + \epsilon^2 G^{(1)}_{\mu\nu}[h^{(2)}] \\+ \epsilon^2 G^{(2)}_{\mu\nu}[h^{(1)},h^{(1)}] + \mathcal{O}(\epsilon^3).
\end{aligned}
\end{equation}
Here, $G^{(1)}_{\mu\nu}$ is the linearized Einstein tensor, and $G^{(2)}_{\mu\nu}$ contains terms quadratic in the perturbations. The first-order field satisfies the linearized vacuum equation
\begin{equation}
\label{eq:o1einstein}
    G^{(1)}_{\mu\nu}[h^{(1)}] = 0,
\end{equation}
while the second-order field is driven by an effective source derived from the self-interaction of $h^{(1)}$:
\begin{equation}
\label{eq:o2einstein}
    G^{(1)}_{\mu\nu}[h^{(2)}] = -G^{(2)}_{\mu\nu}[h^{(1)},h^{(1)}].
\end{equation}

Under appropriate boundary and gauge conditions, solutions to Eqs.\kern0.3em(\ref{eq:o1einstein}) and (\ref{eq:o2einstein}) can be approximated by sums over QNMs and QQNMs, respectively, along with their associated spin-weighted angular functions.
The QQNMs in second-order metric perturbations $h^{(2)}$ are entirely sourced by first-order quantities. Therefore, for a given QQNM, the frequency and amplitude of the second-order offspring component, $\omega^{(2)}$ and $A^{(2)}$, in terms of the corresponding parent QNM frequencies and amplitudes, $\omega^{(1)}$ and $A^{(1)}$, satisfy:\footnote{A given offspring QQNM receives contributions from two parent QNMs and their corresponding mirror modes. Here we temporarily neglect the $m<0$ modes, since the 
BBH simulations considered in this work are all nonprecessing and quasi-circular, for which a reflection symmetry relates each mode to its mirror counterpart.}\kern0.3em\cite{Nakano2007,qqnmfreq1,qqnmfreq2,qqnmfreq3,London:2014}
\begin{equation}
    \begin{aligned}
    \omega^{(2)}_{l_1m_1n_1\times l_2m_2n_2} &= \omega^{(1)}_{l_1m_1n_1}+\omega^{(1)}_{l_2m_2n_2};\\
    A^{(2)}_{l_1m_1n_1\times l_2m_2n_2} &= \mathcal{R}_{l_1m_1n_1\times l_2m_2n_2} A^{(1)}_{l_1m_1n_1} A^{(1)}_{l_2m_2n_2},
    \end{aligned}
\end{equation}
where the triplet $(l,m,n)$ labels a linear QNM, the paired index $(l_1,m_1,n_1)\times(l_2,m_2,n_2)$ labels the corresponding quadratic mode generated by their coupling, and the coefficient $\mathcal R$ is the first-to-second amplitude ratio. 

The ratio $\mathcal{R}$ is a key quantity in second-order BHPT for source-driven perturbations. It depends on the Kerr black hole’s dimensionless spin and on the angular mode under consideration. The self-coupling of the $(2,2,0)$ mode has been studied extensively from multiple perspectives. While the analysis of amplitude ratios originated with Nakano et al.~\cite{Nakano2007,nakano2} in Schwarzschild spacetime, a recent surge of interest was sparked when Mitman~\cite{Mitman2022} and Cheung~\cite{Cheung:2022} extracted QQNMs and $\mathcal R$ using SXS data. Subsequent numerical and theoretical studies~\cite{Yuste:2024,Cheung:2024,Ma:2024,Zhu:2024,Buccio:2024}, however, reported discrepant results. These contradictions were later rationalized by Bourg et al.~\cite{Bourg:2025} via the parity ratio of linear perturbations. Most recently, Khera et al.~\cite{Khera:2025} resolved the remaining inconsistencies by comprehensively analyzing the coupling channels, confirming that the $\mathcal R$ is determined solely by the dimensionless spin.

Taken together, these results establish that $\mathcal R$ depends only on intrinsic black-hole parameters, which motivates the QQNM--memory connection in ringdown explored in this work. 
Since the linear $(2,2,0)$ mode typically dominates the ringdown, this work focuses on the quadratic response it generates, whose leading contribution resides in the $l=m=4$ harmonic.

\subsection{Memory effects}
\label{sec:IIB}
A confirmed detection of the GW memory effect would provide a direct observational probe of nonlinear aspects of general relativity. In particular, it would demonstrate that the radiative degrees of freedom of gravity can leave a permanent imprint on the spacetime metric at null infinity through the fluxes they generate\kern0.3em\cite{Christ:1991,Wiseman:1991,Favata2010MemoryReview}. Such a measurement would also offer an observable handle on the connection between the Bondi–Metzner–Sachs (BMS) asymptotic symmetries and soft-graviton theorems, providing potential empirical access to the infrared structure of gravity\kern0.3em\cite{StromingerZhiboedov2016,Strominger2017LecturesIR}. Because the memory amplitude effectively encodes integrated, anisotropic radiative fluxes, its measurement would provide an independent diagnostic of the total emitted radiation, higher multipoles, and large-scale boundary conditions in compact binary coalescences\kern0.3em\cite{Favata2009NonlinearBBH,Favata2009Revisited,Lasky:2016}.

The Christodoulou nonlinear memory strain of GWs can be expressed as an integral of the GW flux quadrupole moment
\begin{equation}
\delta h^{TT}_{jk}(\tau_R, \Omega)
= \frac{4G}{D c^4}
\int_{-\infty}^{\tau_R} du
\int_{S^2} d\Omega'\,
\frac{dE}{du\, d\Omega'}
\left[
\frac{n_j n_k}{1 - n^l N_l}
\right]^{TT},
\label{eq:memory}
\end{equation}
where $n(\Omega')$ is a unit vector, $N (\Omega)$ is the unit line-of-sight vector drawn from the observer at Earth to the source, and $u$ denotes the retarded time. The energy flux is
\begin{equation}
\label{eq:memory_flux}
    \frac{dE}{du\,d\Omega}=\frac{D^2 c^3}{16\pi G}|\dot h (u,\Omega)|^2,
\end{equation}
where $\dot h \equiv dh/du$ and $h$ is the GW strain. The angular variables are denoted as $\Omega = (\theta, \phi)$, representing the inclination and a reference phase of the source — typically the phase at coalescence in compact-binary systems. The symbol $\tau_R$ refers to the retarded time. The coordinates $\Omega'$ specify a direction on a sphere of radius $D$ centered on the source, where $D$ is the separation between the source and the observer. The label ``\textit{TT}" indicates that the expression is written in the transverse--traceless gauge.

The complex strain $h$ can be written in the polarization basis and expanded in spin-weighted spherical harmonics. 
With $h^{TT}_{jk}$ contracted against the polarization tensors, this gives
\begin{equation}
\label{eq:memory_decompose}
    \begin{aligned}
        h &= h_+ - i h_\times\\
        &=\frac{1}{2}h^{TT}_{jk}(e^{jk}_+ - i e^{jk}_{\times})\\
        &=\sum^{\infty}_{l=2}\sum^{l}_{m=-l} h_{lm}(u)\,{}_{-2}Y_{lm}(\Omega).
    \end{aligned}
\end{equation}
It is convenient to analyze mode couplings in the harmonic basis. First, we substitute the spherical-harmonic expansion in Eq.~(\ref{eq:memory_decompose}) into Eq.~(\ref{eq:memory_flux}), insert the result into Eq.~(\ref{eq:memory}), and contract with the polarization tensors to obtain $\delta h$. 
Projecting $\delta h$ back onto the spherical-harmonic basis then yields
\begin{equation}
\label{eq:memory_main}
    \delta h _{lm} = \frac{D}{4 \pi c }\Gamma^{l_1 m_1 l_2 m_2}_{lm} H_{l_1 m_1 l_2 m_2}(\tau_0,\tau_R).
\end{equation}
In Eq.~(\ref{eq:memory_main}), the quantities $H$ and $\Gamma$ are defined as
\begin{equation}
    H_{l_1 m_1 l_2 m_2}(\tau_0,\tau_R) = \int^{\tau_R}_{\tau_0} du \dot{h}_{l_1m_1}(u) \dot{\bar{h}}_{l_2m_2}(u),
\end{equation}
and
\begin{equation}
    \begin{aligned}
        &\Gamma^{l_1m_1l_2m_2}_{l\,m}\equiv 2\pi \int^1_{-1}
        d(\cos\theta)\,{}_{-2}\bar{Y}_{l\,m}(\theta,0)(e^{jk}_+ - ie^{jk}_\times)
        \\
       &\times\int_{S^2}d\Omega '\,{}_{-2}Y_{l_1m_1}(\Omega')\,{}_{-2}\bar{Y}_{l_2m_2}(\Omega ')\left[\frac{n_jn_k}{1-{n^iN_i}
        } \right]^{TT},
    \end{aligned}
\end{equation}
where an overbar denotes complex conjugation, and the azimuthal index $m$ of the memory mode is fixed by $m = m_1 - m_2$. 
The quantity $H_{l_1 m_1 l_2 m_2}$ is the central object we compute: it represents the time-integrated flux at null infinity through quadratic products of the Bondi news.
The coefficient $\Gamma^{l_1m_1l_2m_2}_{lm}$ is purely geometric and acts as a harmonic-coupling kernel that maps the oscillatory harmonic components $(l_1,m_1)$ and $(l_2,m_2)$ of the strain to the memory component $(l,m)$. In this work, we primarily compute the memory strain using Eq.~(\ref{eq:memory_main}).

\subsection{Bridge coefficients of the nonlinearities}
\label{sec:IIC}
For simplicity, we only present the derivation of one harmonic mode here, but the relationship is certainly general. We utilize the ringdown waveform model purely involving linear QNMs
\begin{equation}
\label{eq:hlinear}
\begin{aligned}
    h^L_{lm}(u)&=\sum^{N-1}_{n=0} h_{lmn}(u)\\
    &=\sum^{N-1}_{n=0} A^{(1)}_{lmn}e^{-i \omega_{lmn}(u-u_{\text{start}})+i\phi_{lmn}}.
\end{aligned}
\end{equation}
Here $A^{(1)}_{lmn}$ and $\omega_{lmn}$ are the amplitude and complex frequency of the linear QNM labeled by $(l,m,n)$, with $\phi_{lmn}$ a constant phase and $u_{\text{start}}$ the chosen ringdown start time.
The integer $N$ specifies how many modes are included in our model, with $N-1$ denoting the highest overtone index. The reference time $u_{\text{start}}$ corresponds to the ringdown starting time, for convenience, we shift the time coordinate so that $u_{\text{start}}=0$. For clarity of presentation, we absorb the spheroidal–spherical harmonic mixing coefficients into the amplitudes.

We consider two arbitrary harmonic components $h_{l_1 m_1}$ and $h_{l_2 m_2}$. 
For notational convenience, we relabel them by $\alpha\equiv(l_1,m_1)$ and $\beta\equiv(l_2,m_2)$. 
Each component is well modeled as a sum of damped sinusoids, so that quadratic products of their time derivatives are integrable on $u\in[0,+\infty)$ and yield finite constants. 
This property enables a direct bridge between QQNMs and the associated nonlinear memory. 
The central ringdown contribution to the memory kernel is
\begin{equation}
\begin{aligned}
   &H^{\mathrm{rd}}_{\alpha\beta}(0,+\infty)=\int_{0}^{+\infty}\sum_{n=0}^{N-1}\sum_{n'=0}^{N-1}
\dot h_{\alpha n}(u)\,\dot{\bar{h}}_{\beta n'}(u)\,du .
\end{aligned}
\end{equation}
Multiplying the two sums yields $N^2$ terms. Among them, the $N$ “diagonal” terms with the equal modes form complex conjugate pairs, canceling the imaginary phase in the exponent and leaving a purely decaying integral; this produces factors $B^{\mathrm{I}}$ that depend only on the QNM frequencies. The remaining $N(N-1)$ cross terms are less simple, but by pairing them one can similarly obtain factors $B^{\mathrm{II}}$ that depend mainly on the QNM frequencies, modulated by the relative phases. Accordingly, we group the $N^2$ terms into equal modes and unequal modes sets
\begin{widetext}
\begin{align}
\label{eq:main_initial}
H^{\mathrm{rd}}_{\alpha \beta}(0,+\infty) &= \int_{0}^{+\infty} \sum_{n,n'}
\Bigg[
\delta_{\alpha n,\beta n'}\,
\dot h_{\alpha n}\,\dot{\bar{h}}_{\beta n'}
+\frac{1-\delta_{\alpha n,\beta n'}}{2}
\Big(
\dot h_{\alpha n}\,\dot{\bar{h}}_{\beta n'}
+\dot h_{\beta n'}\,\dot{\bar{h}}_{\alpha n}
\Big)
\Bigg]\,du \nonumber\\
&=\sum_{n,n'} A^{(1)}_{\alpha n}A^{(1)}_{\beta n'}
\Bigl[
\delta_{\alpha n,\beta n'}\,B^{\rm I}_{\alpha n,\alpha n}
+(1-\delta_{\alpha n,\beta n'})\,B^{\rm II}_{\alpha n,\beta n'}
\Bigr].
\end{align}
\end{widetext}
The factors $B^\mathrm{I}$ satisfy
\begin{equation}
\label{eq:B1}
B^{\rm I}_{\alpha n,\alpha n}:=-\frac{|\omega_{\alpha n}|^{2}}{2\,\Im(\omega_{\alpha n})}.
\end{equation}
Under the QNM convention $\Im(\omega_{\alpha n})<0$, this factor is positive.
The factors $B^\mathrm{II}$ satisfy
\begin{equation}
\label{eq:B2}
B^{\rm II}_{\alpha n,\beta n'}
:=\frac{|\omega_{\alpha n}\,\bar\omega_{\beta n'}|}
      {|\omega_{\alpha n}-\bar\omega_{\beta n'}|}
\cos\,\bigl(\Delta \phi_{\alpha n, \beta n'}+\psi_{\alpha n, \beta n'}\bigr).
\end{equation}
A detailed derivation of Eqs.~(\ref{eq:main_initial})--(\ref{eq:B2}) is given in Appendix~\ref{app:A}. We indicate the real and imaginary parts of a QNM frequency via $\Re(\omega)$ and $\Im(\omega)$. 

\begin{figure}[htbp] 
    \centering
    \includegraphics[width=0.45\textwidth]{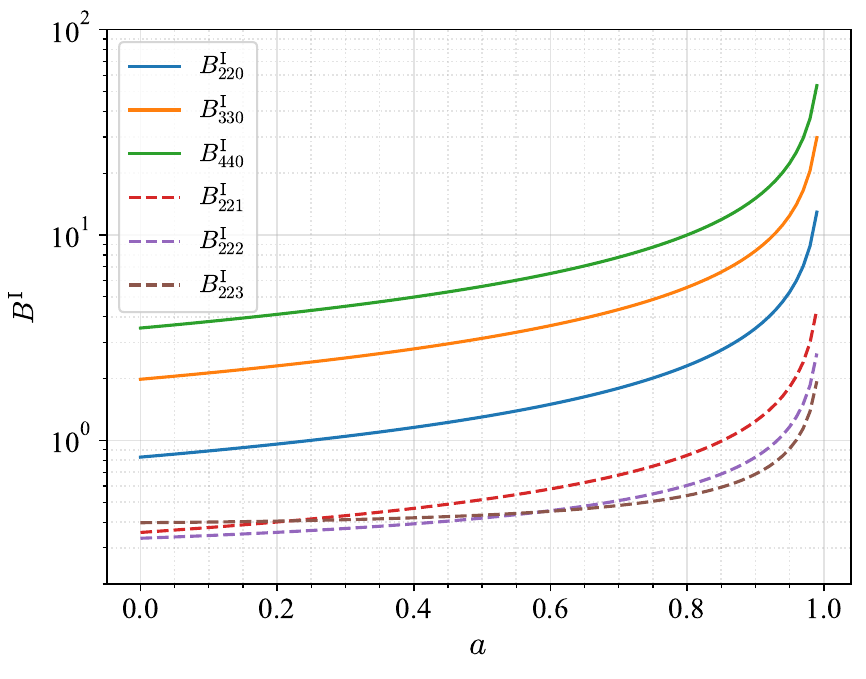} 
    \caption{Dimensionless spin dependence of $B^{\mathrm{I}}$ from QNM self-coupling.} 
    \label{fig:Mval} 
\end{figure}

\begin{figure}[htbp] 
    \centering
    \includegraphics[width=0.45\textwidth]{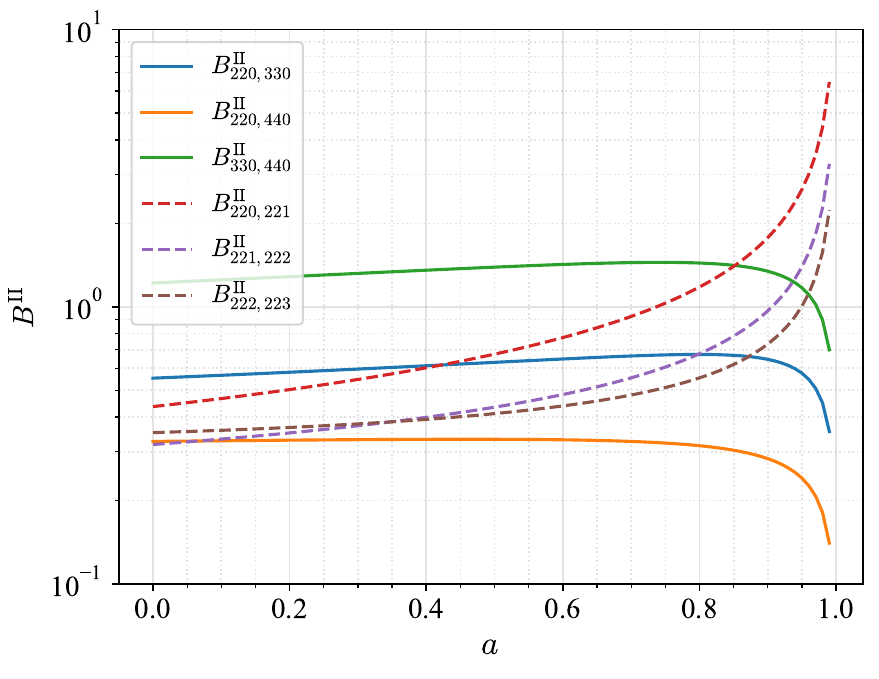} 
    \caption{Dimensionless spin dependence of $B^{\mathrm{II}}$ from QNM cross-coupling.} 
    \label{fig:Nval} 
\end{figure}

Here, in the unit-mass convention, we compute the dependence of $B^{\mathrm{I}}$ and $B^{\mathrm{II}}$ on the dimensionless spin, as shown in Figure\kern0.3em\ref{fig:Mval} and Figure\kern0.3em\ref{fig:Nval}. Figure\kern0.3em\ref{fig:Mval} presents the self-coupling of the fundamental modes for different spheroidal harmonic components, as well as the overtone self-coupling for the $l = m = 2$ component.
Figure\kern0.3em\ref{fig:Nval} shows the cross-couplings among different higher-order spheroidal harmonic components and among different overtones for $\Delta \phi = 0$.
The self-coupling coefficient $B^{\mathrm{I}}$ increases with spin and rises steeply as $a \to 1$. By contrast, $B^{\mathrm{II}}$ shows a weaker spin dependence overall, while the fundamental cross-couplings among spherical-harmonic components are suppressed at high spin.

As recent works have mentioned \cite{Yuste:2024,Cheung:2024,Ma:2024,Zhu:2024,Buccio:2024,Khera:2025}, the amplitude ratio between a QQNM and the product of the matching source-QNM amplitudes depends only on the geometric structure of the remnant black hole, and the ratio can be theoretically determined. We replace amplitudes of linear modes in $H_{\alpha \beta}$ with amplitudes of QQNMs
\begin{equation}
\begin{aligned}
\label{eq:main}
    &\,\,\,\,\,\,\,\,H^{\text{rd}}_{\alpha \beta}(0,+\infty)\\
    &=\sum_{n,n'} \frac{A^{(2)}_{\alpha n \times \beta n'}}{\mathcal R_{\alpha n \times \beta n'}}
    \Bigl[ \delta_{\alpha n,\beta n'}\,B^{\rm I}_{\alpha n,\alpha n} +(1-\delta_{\alpha n,\beta n'})\,B^{\rm II}_{\alpha n,\beta n'} \Bigr]\\
    &=\sum_{n,n'}A^{(2)}_{\alpha n \times \beta n'}\, \Lambda_{\alpha n,\beta n'}.
\end{aligned}
\end{equation} 
The factor $B^\sigma/ \mathcal R$ acts as a bridge between the accumulated memory strain from QNMs and QQNM amplitudes in ringdown. As a result, we define the bridge coefficient $\Lambda$ as
\begin{equation}
\Lambda_{\alpha n,\beta n'}=\frac{B^{\sigma}_{\alpha n,\beta n'}}{\mathcal R_{\alpha n\times \beta n'}},
\qquad
\sigma=
\begin{cases}
\mathrm{I}, & \alpha n=\beta n',\\
\mathrm{II}, & \alpha n\neq \beta n'.
\end{cases}
\end{equation}

Consequently, in the ringdown phase, the accumulated memory strain sourced by the linear modes can be expressed as a sum of QQNM amplitudes weighted by the bridge coefficients $\Lambda$. Since $\Lambda$ depends on the remnant black hole’s mass and spin, the total memory predicted by general relativity follows directly from Eq.~(\ref{eq:main}) once the QQNM amplitudes are measured. This relation provides a new avenue for probing nonlinear gravitational dynamics and for testing gravity.

\section{Tests with SXS Waveforms}
\label{sec:verify}

\begin{figure*}[t] 
  \centering
  \includegraphics[width=0.65\textwidth]{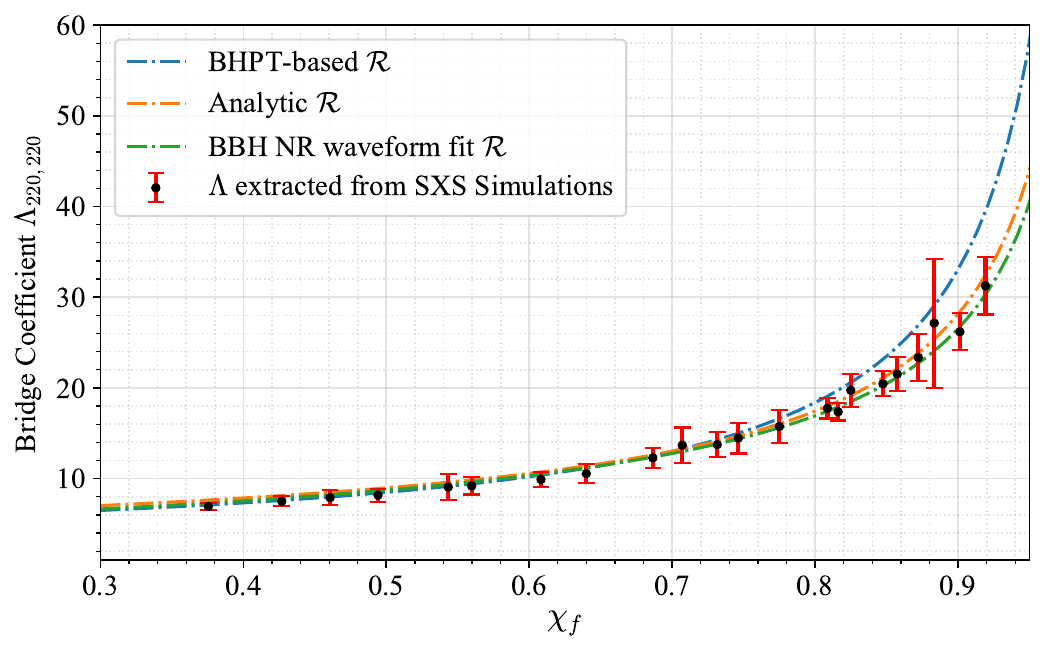} 
  \caption{Verification of the bridge coefficient $\Lambda$ for the $(2,2,0)\times(2,2,0)\rightarrow (4,4)$ coupling channel from the SXS catalog. The horizontal axis shows the dimensionless remnant spin, and the vertical axis shows the value of $\Lambda$ from dimensionless waveforms. Red error bars are $\Lambda$ values measured from a set of SXS BBH simulations spanning the main remnant dimensionless spin range, while the blue~\cite{Yuste:2024}, orange~\cite{Khera:2025}, and green~\cite{Cheung:2024} dash-dotted curves  are theoretical $\Lambda$ predictions obtained from the amplitude ratio $\mathcal R$ using three different methods. This figure confirms the bridge relation between the QQNM amplitude and the full accumulated memory strain in BBH ringdown waveforms.}
  \label{fig:QDM}
\end{figure*}
To provide a precise verification, we analyze a set of 22 numerical simulations of BBH systems (summarized in Table \ref{tab:sxssimulations}), spanning remnant spin $\chi_f \in [0.30,0.92]$. We restrict attention to nearly equal-mass, nonprecessing, quasicircular BBH simulations only. All simulations were performed with the Spectral Einstein Code (\texttt{SpEC}) and are publicly available in the SXS catalog\kern0.3em\cite{sxs1,sxs2,sxs3,sxs4}.
For each dataset, the strain waveform is extracted using Cauchy–characteristic extraction and subsequently transformed into the superrest frame at 250M after $u_{\text{peak}}$\kern0.3em\cite{cce1,cce2,cce3,cce4}, following the procedure of Ref. \cite{cce4} and using the code \texttt{scri} \cite{scri}.

\begin{table}[htbp]
\centering
\caption{\label{tab:sxssimulations}SXS simulation IDs and corresponding remnant dimensionless spin}
\begin{tabular}{c|cccccccc}  
\hline
\hline
ID & 3976 & 3895 & 2083 & 2089 & 2084 & 3533 & 2328 & 2096 \\
$\chi_f$ & 0.376 & 0.427 & 0.461 & 0.494 & 0.543 & 0.560 & 0.608 & 0.640 \\
\hline
ID & 2325 & 3926 & 2097 & 3584 & 2101 & 3921 & 2105 & 3916 \\
$\chi_f$ & 0.686 & 0.707 & 0.732 & 0.746 & 0.775 & 0.809 & 0.816 & 0.825 \\
\hline
ID & 3980 & 2099 & 1495 & 2512 & 3979 & 2423 &  &  \\
$\chi_f$ & 0.847 & 0.857 & 0.872 & 0.883 & 0.901 & 0.919 &  &   \\
\hline
\hline
\end{tabular}
\end{table}
Our procedure for extracting the $\Lambda$ values from the SXS simulations is as follows: we first fit QQNM amplitudes at the corresponding fixed frequencies, and directly extract memory accumulation from linear modes utilizing the same fitting algorithm and \texttt{GWmemory} package\kern0.3em\cite{PhysRevD.98.064031}. 
Finally, the value of $\Lambda$ is obtained by dividing the extracted memory strain by the corresponding QQNM amplitude, after reinstating the geometric coupling factor and the overall prefactors in the memory formula.

The fitting procedure follows the methodology developed in a series of earlier works\kern0.3em\cite{Mitman2022,Cheung:2022,fit1,fit2}. The model we use involves only linear modes and is given by Eq.\kern0.3em(\ref{eq:hlinear}). Meanwhile, the model to extract second order amplitude from $l=m=4$ harmonic is
\begin{equation}
\begin{aligned}
    h^{Q}_{44}(u) = \sum^{N-1}_{n=0}A^{(1)}_{44n}e^{-i \omega _{44n}(u-u_{\text{start}})+i\phi_{44n}} +\\ A_{220\times 220}^{(2)}e^{-i \omega_{220\times220}(u-u_{\text{start}})+i\phi_Q}.
\end{aligned}
\end{equation}
The spherical-spheroidal mixing coefficients are also absorbed in each amplitude.

Given the remnant mass and spin of each simulated event, we use the \texttt{qnm} package\kern0.3em\cite{steinqnm} to compute the QNM frequencies and construct the corresponding basis, and then perform a least-squares fit of the amplitudes. We fix $u_f = 100M$ as the end time of the fitting window. To account for systematic uncertainties in the choice of the starting time, we vary $u_0$ in the range $[15M,28M]$, and quote the average value together with an error bar estimated from this spread. For the $\ell=m=2$ harmonic  we use $N=3$ basis modes in the fit, while for extracting QQNMs in the $\ell=m=4$ harmonic we increase the number of basis modes to improve stability: we take $N=3$ for remnants with 
$a\le 0.7$ and $N=4$ for $a > 0.7$. This choice is well motivated, since larger remnant spins are expected to excite a richer mode content, and in our tests the error bars for $N=4$ are no larger than in the other cases. Moreover, the number of basis modes we use always remains within the analytically allowed upper limit given in Refs.\kern0.3em\cite{fit2,oshita1}. 

Using the above procedure to extract the QNM content, we employ \texttt{GWmemory} to obtain the corresponding memory strain.
By dividing the extracted memory strain by the associated QQNM amplitude, after reinstating the overall prefactors of the full memory expression, we infer the bridge coefficient $\Lambda$ from numerical-relativity simulations of BBH coalescences. Note that we work with the dimensionless strain $rh/M$, so the overall distance coefficient $D$ in Eq.\kern0.3em(\ref{eq:memory_main}) cancels out and need not be included. Our validation focuses primarily on the dominant $(2,2,0)\times(2,2,0)\rightarrow (4,4)$ channel. 

Figure~\ref{fig:QDM} presents our verification results. We have adopted three types of amplitude ratios, including (i) a BHPT-based numerical evolution with Gaussian wave-packet initial data \cite{Yuste:2024}, (ii) a purely analytical ratio that considers the complete set of coupling channels \cite{Khera:2025}, and (iii) a fitted ratio derived from the SXS catalog \cite{Cheung:2024}. The blue, orange, and green dash-dotted curves correspond to these three amplitude ratios, respectively. The black dots with red error bars mark the bridge coefficient $\Lambda$ extracted from the numerical relativity BBH waveforms. It can be observed that the two sets align well with each other.

We show that Eq.\kern0.3em(\ref{eq:main}) holds in the ringdown signal from BBH coalescences, and we validate it using numerical-relativity waveforms. This relation offers a potential new test of gravity with future observations, and it motivates further investigation into the intrinsic connections among different nonlinear effects during BBH coalescence. 

\section{Observational Prospects}
\label{sec:detection}
In the ringdown regime, we establish a bridge relation that links QQNM amplitudes sourced by linear QNMs to the accumulated nonlinear memory. 
Using this relation to test gravity requires a simultaneous measurement of both QQNMs and the memory signal. 
Following the demonstration that QQNMs can contribute non-negligibly to BBH ringdowns~\cite{Mitman2022,Cheung:2022}, several studies have examined their detectability~\cite{Yi:2024,Lagos:2024,Shi:2024,Khera:2025} and have reached broadly optimistic conclusions. 
There is even a recent claim of QQNM evidence in the GWTC--4 event GW250114~\cite{Yang:2025,Wang:2026rev}. 
By contrast, despite extensive searches, there is still no convincing observational evidence for the memory effect. 
This is largely because memory is a direct-current feature with a very small amplitude~\cite{Cheung:2024memory,PhysRevD.101.023011,Lasky:2016,Zhao:2021,Ebersold:2020}. 
Since memory is intrinsically low frequency, future space-based detectors with improved sensitivity in this band should offer a clearer observational window.

In this setting, our bridge relation becomes testable when the signal carries a sufficiently clear memory imprint. 
Reference~\cite{Incha:2024} shows that, for a ringdown with remnant mass $\sim 10^{5.5} M_\odot$ at $z\lesssim 1$, the memory contribution can already achieve a substantial signal-to-noise ratio (SNR) in LISA. 
Because LISA measures time-varying optical path-length differences, the long-time plateau of the accumulated memory is not observed directly. 
Instead, the memory enters the data as a pulse-like contribution concentrated near merger and early ringdown. 
As illustrated in Fig.~2 of Ref.~\cite{Incha:2024}, the ringdown contributes most of the memory SNR, making it the most relevant epoch for confronting our relation with observations.

In real observations, one must address a practical point. Linear QNMs typically dominate the ringdown signal, but the total memory strain in principle also includes contributions from QQNMs. It is therefore important to ask whether the memory from QQNMs can be neglected compared with the memory from QNMs, and under what conditions.

We use multimode ringdown amplitudes fitted to the SXS catalog in Ref.~\cite{Cheung:2024} and the ratios of selected second-order to first-order QNM amplitudes reported in the supplementary materials of Ref.~\cite{Khera:2025}. Since our goal is only an order-of-magnitude comparison, these inputs are sufficient.
We label each quadratic component by its parent pair, $(l_1,m_1,n_1)\times(l_2,m_2,n_2)$, which sets its second-order frequency. We denote the mirror counterpart of \((\ell,m,n)\) by \((\ell,-m,n)^*\), where the star indicates the complex-conjugate branch. The corresponding frequencies satisfy
\(\omega_{\ell m n}=-\omega^*_{\ell,-m,n}\).

We consider four representative offspring components from Ref.~\cite{Khera:2025}, labeled by their parent pairs:
$(2,2,0)\!\times\!(2,2,0)$, $(2,2,0)\!\times\!(3,3,0)$, $(4,4,0)\!\times\!(2,-2,0)^*$, and $(3,3,0)\!\times\!(2,-2,0)^*$.
A given offspring component also receives contributions from the mirror counterparts of its parent pair. For nonprecessing binaries, reflection symmetry makes these contributions redundant, so we use the parent pairs listed above as a convenient reference.

\begin{figure}[t] 

  \centering
  \includegraphics[width=0.45\textwidth]{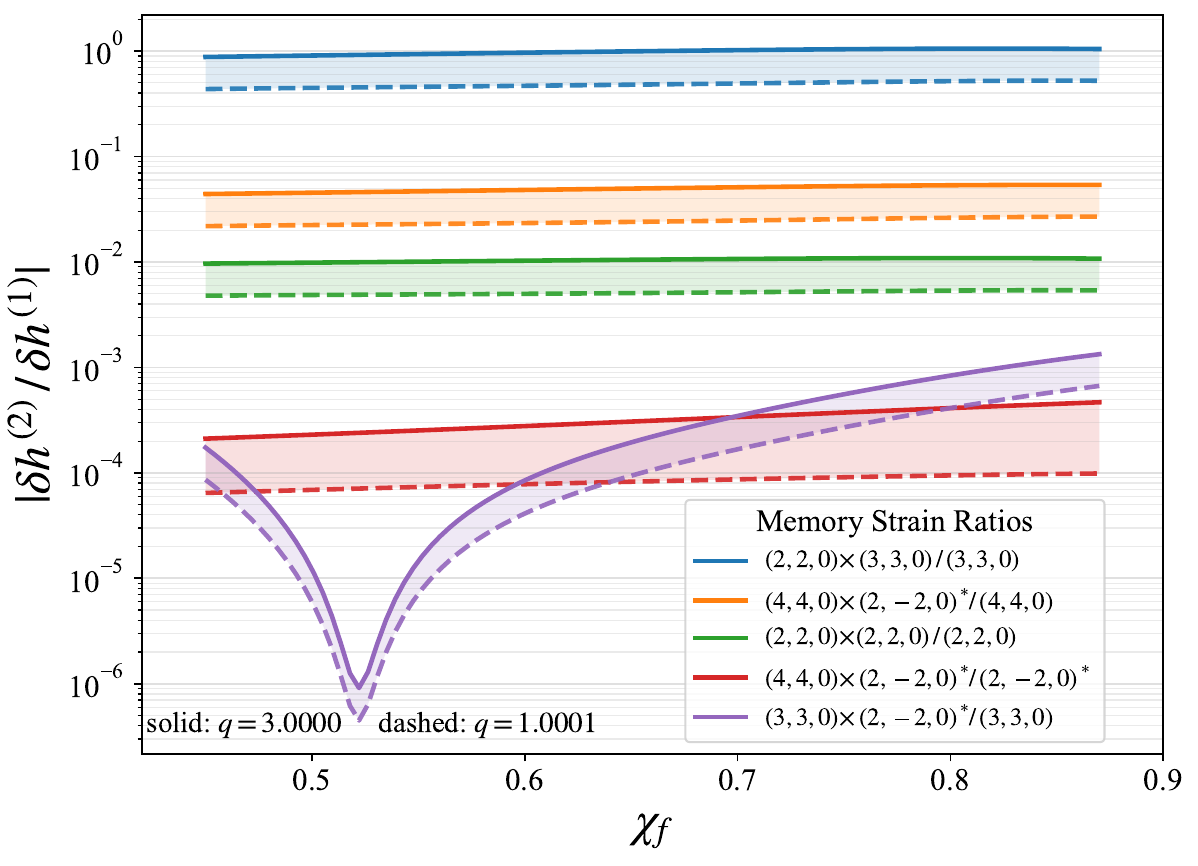} 
  \caption{Ratios of the memory strain sourced by an offspring QQNM to that sourced by a parent QNM, shown as $|\delta h^{(2)}/\delta h^{(1)}|$ versus the remnant spin $\chi_f$.
Solid curves use $q=3.0000$ and dashed curves use $q=1.0001$.
The legend labels each QQNM by its parent pair and specifies the parent mode used in the denominator.
Shaded bands span the two mass ratios for the same component.
Most ratios are below $5\times10^{-2}$ with weak spin dependence.
The largest variation occurs for $(3,3,0)\times(2,-2,0)^{*}$, while $(2,2,0)\times(3,3,0)/(3,3,0)$ can approach unity at high $\chi_f$.}
  \label{fig:MemoryRatios}
\end{figure}

We analyze aligned-spin, nonprecessing BBH simulations at $q=3.0000$ (solid curve) and $q=1.0001$ (dashed curve) and report the resulting memory-accumulation ratios in Figure~\ref{fig:MemoryRatios}. 
Each ratio is defined as $|\delta h^{(2)}/\delta h^{(1)}|$ in the figure, where the parent in the denominator is specified in the legend.
Here $\delta h$ is evaluated at $\theta=\pi/2$ and $\phi=0$.
For the dominant channels, the QQNM-induced memory is typically about two orders of magnitude smaller than the memory from the corresponding parent QNMs. The ratios show only a weak dependence on the remnant spin, with the notable exception of $(3,3,0)\times(2,-2,0)^{*}$. This separation of scales implies that the offspring contributions can be treated as perturbative corrections to the parent-mode memory, so the bridge relation proposed here should be relevant for future space-based detectors.
An exception is $(2,2,0)\times(3,3,0)$: its memory can match or exceed the parent $(3,3,0)$ memory at large $q$ and $\chi_f$, but its absolute contribution remains limited.
These features motivate a more systematic study in the regime of high mass ratio, high spin, and possible precession.

\vspace{1.0\baselineskip}

\section{Discussion and Outlook}
\label{sec:conclusion}
The two nonlinear effects studied here share a common origin in the ringdown. QQNMs capture an oscillatory nonlinear response in the strong-field near-zone, whereas the Christodoulou memory is sourced by the accumulated outgoing energy flux at null infinity. We derive a quantitative relation between them. For any quadratic combination of linear QNMs, the offspring QQNM amplitude is proportional to the total accumulated memory strain, with a bridge coefficient determined primarily by the remnant black hole parameters. Equivalently, the accumulated ringdown memory can be written as a sum of QQNM amplitudes weighted by the corresponding bridge coefficients. We show that this relation is present in BBH ringdowns, enabling a new test of gravity and offering a complementary probe of strong-field nonlinear dynamics. This opens an experimentally accessible route to consistency tests that directly target nonlinear effects.

While current ground-based detectors have modest sensitivity to nonlinear effects, future space-based observatories should yield many more high-SNR events. In particular, sources with remnant mass around $10^{5.5} M_\odot$ at $z \lesssim 1$ are promising targets for such a test~\cite{Incha:2024}. We also find that, for most channels, the memory 
generated by offspring QQNMs is a small correction to the parent QNM contribution, which supports the applicability of our framework. The bridge relation offers a new route to model ringdown memory using higher-order mode amplitudes, which merits further investigation. From a data-analysis perspective, since QQNMs are typically much stronger than memory, they can serve as valuable priors or constraints that inform memory inference. We anticipate further insights into the intrinsic connections among nonlinear observables in the strong-field regime.

\section*{ACKNOWLEDGMENTS}

We thank Zhoujian Cao and Cong-Yuan Yue for helpful discussions. This research is supported in part by the National Key R$\&$D Program of China, grant number 2020YFC2201300, and the National Natural Science Foundation of China, grant numbers 12375058 and 12361141825.

\onecolumngrid
\appendix
\section{Derivation of Eqs.~(\ref{eq:main_initial})--(\ref{eq:B2})}
\label{app:A}

In this Appendix we present a compact derivation of Eqs.~(\ref{eq:main_initial})--(\ref{eq:B2}). 
We adopt the standard QNM convention $\omega_{\alpha n}=\Re(\omega_{\alpha n})+i\,\Im(\omega_{\alpha n})$ with $\Im(\omega_{\alpha n})<0$, which guarantees the convergence of all integrals over $u\in[0,+\infty)$. 
The Kronecker delta $\delta_{\alpha n,\beta n'}$ picks out identical mode labels, while the factor $(1-\delta_{\alpha n,\beta n'})/2$ avoids double counting in the sum over unordered mode pairs in the cross term. For convenience, we decompose Eq.~(\ref{eq:main_initial}) into diagonal (self-coupling) and off-diagonal (cross-coupling) parts:
\begin{equation}
    H^{rd}_{\alpha \beta}(0,+\infty)=
    \sum_{nn'}\big(\delta_{\alpha n,\beta n'}I^{\textrm{I}}_{\alpha n,\alpha n} + \frac{1-\delta_{\alpha n,\beta n'}}{2} I^{\textrm{II}}_{\alpha n,\beta n'} \big).
\end{equation}

The diagonal term $I^{\textrm{I}}_{\alpha n,\alpha n}$ follows directly from the self-coupling terms:
\begin{equation}
    \begin{aligned}
        I^{\textrm{I}}_{\alpha n,\alpha n} &= \int^{+\infty}_0 du \, \dot{h}_{\alpha n} \dot{\bar{h}}_{\alpha n}\\
        &=A^{(1)2}_{\alpha n}\,|\omega_{\alpha n}|^2 \int^{+\infty}_0 du\, e^{2 \Im(\omega_{\alpha n})u}\\
        &=-A^{(1)2}_{\alpha n}\,|\omega_{\alpha n}|^2/2\Im(\omega_{\alpha n}).
    \end{aligned}
\end{equation}
We then introduce the shorthand
\begin{equation}
    B^\textrm{I}_{\alpha n,\alpha n} = -\frac{|\omega_{\alpha n}|^2}{2\Im(\omega_{\alpha n})},
\end{equation}
which isolates the frequency-dependent factor appearing in the self-coupling contribution.
It is positive under the convention $\Im(\omega_{\alpha n})<0$.

The off-diagonal term $I^{\textrm{II}}_{\alpha n,\beta n'}$ can be treated analogously, while keeping track of relative phases. 
In going to the second line, we use the conjugation $\dot h_{\beta n'}\dot{\bar h}_{\alpha n}=\overline{\dot h_{\alpha n}\dot{\bar h}_{\beta n'}}$.
In the last line, we rewrite the resulting real part as a cosine of the total phase $\Delta \phi_{\alpha n, \beta n'}+ \psi_{\alpha n, \beta n'}$, obtaining a compact amplitude--phase representation:
\begin{equation}
    \begin{aligned}
        I^{\textrm{II}}_{\alpha n,\beta n'} &= \int^{+\infty}_0 du\Bigl(\dot h_{\alpha n}\dot{\bar h}_{\beta n'}+\dot h_{\beta n'}\dot{\bar h}_{\alpha n}\Bigr)\\
        &=\int^{+\infty}_0\, du \, 2\Re\!\left(\dot h_{\alpha n}\dot{\bar h}_{\beta n'}\right)\\
        &=2A^{(1)}_{\alpha n}A^{(1)}_{\beta n'}
        \Re\left(\omega_{\alpha n} \bar{\omega}_{\beta n'} e^{i(\phi_{\alpha n}-\phi_{\beta n'})}
        \int^{+\infty}_0\, du\, e^{-i(\omega_{\alpha n} - \bar{\omega}_{\beta n'})u}\right)\\
        &=2A^{(1)}_{\alpha n}A^{(1)}_{\beta n'}\Re\left[ \frac{\omega_{\alpha n} \bar{\omega}_{\beta n'}}{-\Im(\omega_{\alpha n} - \bar{\omega}_{\beta n'}) + i \Re(\omega_{\alpha n} - \bar{\omega}_{\beta n'})} e^{i(\phi_{\alpha n}-\phi_{\beta n'})}   \right]\\
        &=2A^{(1)}_{\alpha n}A^{(1)}_{\beta n'} \frac{|\omega_{\alpha n}\,\bar\omega_{\beta n'}|}
        {|\omega_{\alpha n}-\bar\omega_{\beta n'}|}
        \cos\bigl(\Delta \phi_{\alpha n, \beta n'}+ \psi_{\alpha n, \beta n'}\bigr).
    \end{aligned}
\end{equation}
Here the phase offset $\psi$ is defined by
\begin{equation}
    \psi_{\alpha n, \beta n'} = \arg \omega_{\alpha n} + \arg \bar{\omega}_{\beta n'} - \arg (\omega_{\alpha n}-\bar\omega_{\beta n'})-\frac{\pi}{2},
\end{equation}
and $\Delta \phi_{\alpha n, \beta n'} = \phi_{\alpha n}-\phi_{\beta n'}$, so that the cross term is expressed as the product of the mode amplitudes and a frequency-dependent coefficient times a cosine of the total phase. 
Accordingly, we define
\begin{equation}
    B^\textrm{II}_{\alpha n, \beta n'}=\frac{|\omega_{\alpha n}\,\bar\omega_{\beta n'}|}
        {|\omega_{\alpha n}-\bar\omega_{\beta n'}|}
        \cos\bigl(\Delta \phi_{\alpha n, \beta n'}+ \psi_{\alpha n, \beta n'}\bigr).
\end{equation}
These expressions provide exactly the structure needed to rewrite Eq.~(\ref{eq:main_initial}) in the compact form reported in Eqs.~(\ref{eq:main_initial})--(\ref{eq:B2}).

\twocolumngrid

\newpage
\bibliographystyle{apsrev4-2}

\begin{thebibliography}{61}%
\makeatletter
\providecommand \@ifxundefined [1]{%
 \@ifx{#1\undefined}
}%
\providecommand \@ifnum [1]{%
 \ifnum #1\expandafter \@firstoftwo
 \else \expandafter \@secondoftwo
 \fi
}%
\providecommand \@ifx [1]{%
 \ifx #1\expandafter \@firstoftwo
 \else \expandafter \@secondoftwo
 \fi
}%
\providecommand \natexlab [1]{#1}%
\providecommand \enquote  [1]{``#1''}%
\providecommand \bibnamefont  [1]{#1}%
\providecommand \bibfnamefont [1]{#1}%
\providecommand \citenamefont [1]{#1}%
\providecommand \href@noop [0]{\@secondoftwo}%
\providecommand \href [0]{\begingroup \@sanitize@url \@href}%
\providecommand \@href[1]{\@@startlink{#1}\@@href}%
\providecommand \@@href[1]{\endgroup#1\@@endlink}%
\providecommand \@sanitize@url [0]{\catcode `\\12\catcode `\$12\catcode
  `\&12\catcode `\#12\catcode `\^12\catcode `\_12\catcode `\%12\relax}%
\providecommand \@@startlink[1]{}%
\providecommand \@@endlink[0]{}%
\providecommand \url  [0]{\begingroup\@sanitize@url \@url }%
\providecommand \@url [1]{\endgroup\@href {#1}{\urlprefix }}%
\providecommand \urlprefix  [0]{URL }%
\providecommand \Eprint [0]{\href }%
\providecommand \doibase [0]{https://doi.org/}%
\providecommand \selectlanguage [0]{\@gobble}%
\providecommand \bibinfo  [0]{\@secondoftwo}%
\providecommand \bibfield  [0]{\@secondoftwo}%
\providecommand \translation [1]{[#1]}%
\providecommand \BibitemOpen [0]{}%
\providecommand \bibitemStop [0]{}%
\providecommand \bibitemNoStop [0]{.\EOS\space}%
\providecommand \EOS [0]{\spacefactor3000\relax}%
\providecommand \BibitemShut  [1]{\csname bibitem#1\endcsname}%
\let\auto@bib@innerbib\@empty
\bibitem [{\citenamefont {Isi}\ \emph {et~al.}(2019)\citenamefont {Isi},
  \citenamefont {Giesler}, \citenamefont {Farr}, \citenamefont {Scheel},\ and\
  \citenamefont {Teukolsky}}]{test:rd1}%
  \BibitemOpen
  \bibfield  {author} {\bibinfo {author} {\bibfnamefont {M.}~\bibnamefont
  {Isi}}, \bibinfo {author} {\bibfnamefont {M.}~\bibnamefont {Giesler}},
  \bibinfo {author} {\bibfnamefont {W.~M.}\ \bibnamefont {Farr}}, \bibinfo
  {author} {\bibfnamefont {M.~A.}\ \bibnamefont {Scheel}},\ and\ \bibinfo
  {author} {\bibfnamefont {S.~A.}\ \bibnamefont {Teukolsky}},\ }\href
  {https://doi.org/10.1103/PhysRevLett.123.111102} {\bibfield  {journal}
  {\bibinfo  {journal} {Phys. Rev. Lett.}\ }\textbf {\bibinfo {volume} {123}},\
  \bibinfo {pages} {111102} (\bibinfo {year} {2019})},\ \Eprint
  {https://arxiv.org/abs/1905.00869} {arXiv:1905.00869 [gr-qc]} \BibitemShut
  {NoStop}%
\bibitem [{\citenamefont {Abbott}\ \emph {et~al.}(2021)\citenamefont {Abbott}
  \emph {et~al.}}]{test:rd2}%
  \BibitemOpen
  \bibfield  {author} {\bibinfo {author} {\bibfnamefont {R.}~\bibnamefont
  {Abbott}} \emph {et~al.} (\bibinfo {collaboration} {LIGO Scientific
  Collaboration and Virgo Collaboration and KAGRA Collaboration}),\ }\href@noop
  {} {\  (\bibinfo {year} {2021})},\ \Eprint {https://arxiv.org/abs/2112.06861}
  {arXiv:2112.06861 [gr-qc]} \BibitemShut {NoStop}%
\bibitem [{\citenamefont {Cardoso}\ \emph {et~al.}(2019)\citenamefont
  {Cardoso}, \citenamefont {Kimura}, \citenamefont {Maselli},\ and\
  \citenamefont {Berti}}]{test:rd3}%
  \BibitemOpen
  \bibfield  {author} {\bibinfo {author} {\bibfnamefont {V.}~\bibnamefont
  {Cardoso}}, \bibinfo {author} {\bibfnamefont {M.}~\bibnamefont {Kimura}},
  \bibinfo {author} {\bibfnamefont {A.}~\bibnamefont {Maselli}},\ and\ \bibinfo
  {author} {\bibfnamefont {E.}~\bibnamefont {Berti}},\ }\href
  {https://doi.org/10.1103/PhysRevD.99.104077} {\bibfield  {journal} {\bibinfo
  {journal} {Phys. Rev. D}\ }\textbf {\bibinfo {volume} {99}},\ \bibinfo
  {pages} {104077} (\bibinfo {year} {2019})}\BibitemShut {NoStop}%
\bibitem [{\citenamefont {Berti}\ \emph {et~al.}(2025)\citenamefont {Berti},
  \citenamefont {Cardoso}, \citenamefont {Carullo} \emph {et~al.}}]{test:rd4}%
  \BibitemOpen
  \bibfield  {author} {\bibinfo {author} {\bibfnamefont {E.}~\bibnamefont
  {Berti}}, \bibinfo {author} {\bibfnamefont {V.}~\bibnamefont {Cardoso}},
  \bibinfo {author} {\bibfnamefont {G.}~\bibnamefont {Carullo}}, \emph
  {et~al.},\ }\href@noop {} {\  (\bibinfo {year} {2025})},\ \Eprint
  {https://arxiv.org/abs/2505.23895} {arXiv:2505.23895 [gr-qc]} \BibitemShut
  {NoStop}%
\bibitem [{\citenamefont {Chung}\ and\ \citenamefont {Yunes}(2025)}]{test:rd5}%
  \BibitemOpen
  \bibfield  {author} {\bibinfo {author} {\bibfnamefont {A.~K.-W.}\
  \bibnamefont {Chung}}\ and\ \bibinfo {author} {\bibfnamefont
  {N.}~\bibnamefont {Yunes}},\ }\href@noop {} {\  (\bibinfo {year} {2025})},\
  \Eprint {https://arxiv.org/abs/2506.14695} {arXiv:2506.14695 [gr-qc]}
  \BibitemShut {NoStop}%
\bibitem [{\citenamefont {Ota}\ and\ \citenamefont
  {Chirenti}(2020)}]{test:rd6}%
  \BibitemOpen
  \bibfield  {author} {\bibinfo {author} {\bibfnamefont {I.}~\bibnamefont
  {Ota}}\ and\ \bibinfo {author} {\bibfnamefont {C.}~\bibnamefont {Chirenti}},\
  }\href {https://doi.org/10.1103/PhysRevD.101.104005} {\bibfield  {journal}
  {\bibinfo  {journal} {Phys. Rev. D}\ }\textbf {\bibinfo {volume} {101}},\
  \bibinfo {pages} {104005} (\bibinfo {year} {2020})}\BibitemShut {NoStop}%
\bibitem [{\citenamefont {Nakano}\ and\ \citenamefont
  {Ioka}(2007)}]{Nakano2007}%
  \BibitemOpen
  \bibfield  {author} {\bibinfo {author} {\bibfnamefont {H.}~\bibnamefont
  {Nakano}}\ and\ \bibinfo {author} {\bibfnamefont {K.}~\bibnamefont {Ioka}},\
  }\href {https://doi.org/10.1103/PhysRevD.76.084007} {\bibfield  {journal}
  {\bibinfo  {journal} {Phys. Rev. D}\ }\textbf {\bibinfo {volume} {76}},\
  \bibinfo {pages} {084007} (\bibinfo {year} {2007})},\ \Eprint
  {https://arxiv.org/abs/0708.0450} {arXiv:0708.0450 [gr-qc]} \BibitemShut
  {NoStop}%
\bibitem [{\citenamefont {Ioka}\ and\ \citenamefont {Nakano}(2007)}]{nakano2}%
  \BibitemOpen
  \bibfield  {author} {\bibinfo {author} {\bibfnamefont {K.}~\bibnamefont
  {Ioka}}\ and\ \bibinfo {author} {\bibfnamefont {H.}~\bibnamefont {Nakano}},\
  }\href {https://doi.org/10.1103/PhysRevD.76.061503} {\bibfield  {journal}
  {\bibinfo  {journal} {Phys. Rev. D}\ }\textbf {\bibinfo {volume} {76}},\
  \bibinfo {pages} {061503} (\bibinfo {year} {2007})}\BibitemShut {NoStop}%
\bibitem [{\citenamefont {Mitman}\ \emph {et~al.}(2023)\citenamefont {Mitman}
  \emph {et~al.}}]{Mitman2022}%
  \BibitemOpen
  \bibfield  {author} {\bibinfo {author} {\bibfnamefont {K.}~\bibnamefont
  {Mitman}} \emph {et~al.},\ }\href
  {https://doi.org/10.1103/PhysRevLett.130.081402} {\bibfield  {journal}
  {\bibinfo  {journal} {Phys. Rev. Lett.}\ }\textbf {\bibinfo {volume} {130}},\
  \bibinfo {pages} {081402} (\bibinfo {year} {2023})},\ \Eprint
  {https://arxiv.org/abs/2208.07380} {arXiv:2208.07380 [gr-qc]} \BibitemShut
  {NoStop}%
\bibitem [{\citenamefont {Cheung}\ \emph {et~al.}(2023)\citenamefont {Cheung}
  \emph {et~al.}}]{Cheung:2022}%
  \BibitemOpen
  \bibfield  {author} {\bibinfo {author} {\bibfnamefont {M.~H.-Y.}\
  \bibnamefont {Cheung}} \emph {et~al.},\ }\href
  {https://doi.org/10.1103/PhysRevLett.130.081401} {\bibfield  {journal}
  {\bibinfo  {journal} {Phys. Rev. Lett.}\ }\textbf {\bibinfo {volume} {130}},\
  \bibinfo {pages} {081401} (\bibinfo {year} {2023})},\ \Eprint
  {https://arxiv.org/abs/2208.07374} {arXiv:2208.07374 [gr-qc]} \BibitemShut
  {NoStop}%
\bibitem [{\citenamefont {Christodoulou}(1991)}]{Christ:1991}%
  \BibitemOpen
  \bibfield  {author} {\bibinfo {author} {\bibfnamefont {D.}~\bibnamefont
  {Christodoulou}},\ }\href {https://doi.org/10.1103/PhysRevLett.67.1486}
  {\bibfield  {journal} {\bibinfo  {journal} {Physical Review Letters}\
  }\textbf {\bibinfo {volume} {67}},\ \bibinfo {pages} {1486} (\bibinfo {year}
  {1991})}\BibitemShut {NoStop}%
\bibitem [{\citenamefont {Wiseman}\ and\ \citenamefont
  {Will}(1991)}]{Wiseman:1991}%
  \BibitemOpen
  \bibfield  {author} {\bibinfo {author} {\bibfnamefont {A.~G.}\ \bibnamefont
  {Wiseman}}\ and\ \bibinfo {author} {\bibfnamefont {C.~M.}\ \bibnamefont
  {Will}},\ }\href {https://doi.org/10.1103/PhysRevD.44.R2945} {\bibfield
  {journal} {\bibinfo  {journal} {Physical Review D}\ }\textbf {\bibinfo
  {volume} {44}},\ \bibinfo {pages} {R2945} (\bibinfo {year}
  {1991})}\BibitemShut {NoStop}%
\bibitem [{\citenamefont {Favata}(2010)}]{Favata2010MemoryReview}%
  \BibitemOpen
  \bibfield  {author} {\bibinfo {author} {\bibfnamefont {M.}~\bibnamefont
  {Favata}},\ }\href {https://doi.org/10.1088/0264-9381/27/8/084036} {\bibfield
   {journal} {\bibinfo  {journal} {Classical and Quantum Gravity}\ }\textbf
  {\bibinfo {volume} {27}},\ \bibinfo {pages} {084036} (\bibinfo {year}
  {2010})},\ \Eprint {https://arxiv.org/abs/1003.3486} {arXiv:1003.3486
  [gr-qc]} \BibitemShut {NoStop}%
\bibitem [{\citenamefont {Cardoso}\ \emph {et~al.}(2024)\citenamefont
  {Cardoso}, \citenamefont {Carullo}, \citenamefont {De~Amicis}, \citenamefont
  {Duque}, \citenamefont {Katagiri}, \citenamefont {Pereniguez}, \citenamefont
  {Redondo-Yuste}, \citenamefont {Spieksma},\ and\ \citenamefont
  {Zhong}}]{Cardoso2024}%
  \BibitemOpen
  \bibfield  {author} {\bibinfo {author} {\bibfnamefont {V.}~\bibnamefont
  {Cardoso}}, \bibinfo {author} {\bibfnamefont {G.}~\bibnamefont {Carullo}},
  \bibinfo {author} {\bibfnamefont {M.}~\bibnamefont {De~Amicis}}, \bibinfo
  {author} {\bibfnamefont {F.}~\bibnamefont {Duque}}, \bibinfo {author}
  {\bibfnamefont {T.}~\bibnamefont {Katagiri}}, \bibinfo {author}
  {\bibfnamefont {D.}~\bibnamefont {Pereniguez}}, \bibinfo {author}
  {\bibfnamefont {J.}~\bibnamefont {Redondo-Yuste}}, \bibinfo {author}
  {\bibfnamefont {T.~F.~M.}\ \bibnamefont {Spieksma}},\ and\ \bibinfo {author}
  {\bibfnamefont {Z.}~\bibnamefont {Zhong}},\ }\href
  {https://doi.org/10.1103/PhysRevD.109.L121502} {\bibfield  {journal}
  {\bibinfo  {journal} {Phys. Rev. D}\ }\textbf {\bibinfo {volume} {109}},\
  \bibinfo {pages} {L121502} (\bibinfo {year} {2024})},\ \Eprint
  {https://arxiv.org/abs/2405.12290} {arXiv:2405.12290 [gr-qc]} \BibitemShut
  {NoStop}%
\bibitem [{\citenamefont {Ma}\ \emph {et~al.}(2025)\citenamefont {Ma},
  \citenamefont {Scheel}, \citenamefont {Moxon}, \citenamefont {Nelli},
  \citenamefont {Deppe}, \citenamefont {Kidder}, \citenamefont {Throwe},\ and\
  \citenamefont {Vu}}]{Ma2024}%
  \BibitemOpen
  \bibfield  {author} {\bibinfo {author} {\bibfnamefont {S.}~\bibnamefont
  {Ma}}, \bibinfo {author} {\bibfnamefont {M.~A.}\ \bibnamefont {Scheel}},
  \bibinfo {author} {\bibfnamefont {J.}~\bibnamefont {Moxon}}, \bibinfo
  {author} {\bibfnamefont {K.~C.}\ \bibnamefont {Nelli}}, \bibinfo {author}
  {\bibfnamefont {N.}~\bibnamefont {Deppe}}, \bibinfo {author} {\bibfnamefont
  {L.~E.}\ \bibnamefont {Kidder}}, \bibinfo {author} {\bibfnamefont
  {W.}~\bibnamefont {Throwe}},\ and\ \bibinfo {author} {\bibfnamefont {N.~L.}\
  \bibnamefont {Vu}},\ }\href {https://doi.org/10.1103/jd26-8q5w} {\bibfield
  {journal} {\bibinfo  {journal} {Phys. Rev. D}\ }\textbf {\bibinfo {volume}
  {112}},\ \bibinfo {pages} {024003} (\bibinfo {year} {2025})},\ \Eprint
  {https://arxiv.org/abs/2412.06906} {arXiv:2412.06906 [gr-qc]} \BibitemShut
  {NoStop}%
\bibitem [{\citenamefont {Ling}\ \emph {et~al.}(2025)\citenamefont {Ling},
  \citenamefont {Shah},\ and\ \citenamefont {Wong}}]{Ling2025}%
  \BibitemOpen
  \bibfield  {author} {\bibinfo {author} {\bibfnamefont {S.}~\bibnamefont
  {Ling}}, \bibinfo {author} {\bibfnamefont {S.}~\bibnamefont {Shah}},\ and\
  \bibinfo {author} {\bibfnamefont {S.~S.~C.}\ \bibnamefont {Wong}},\ }\href
  {https://doi.org/10.1103/22lc-62gj} {\bibfield  {journal} {\bibinfo
  {journal} {Phys. Rev. D}\ }\textbf {\bibinfo {volume} {112}},\ \bibinfo
  {pages} {024008} (\bibinfo {year} {2025})},\ \Eprint
  {https://arxiv.org/abs/2503.19967} {arXiv:2503.19967 [gr-qc]} \BibitemShut
  {NoStop}%
\bibitem [{\citenamefont {Kehagias}\ and\ \citenamefont
  {Riotto}(2025)}]{Kehagias2025}%
  \BibitemOpen
  \bibfield  {author} {\bibinfo {author} {\bibfnamefont {A.}~\bibnamefont
  {Kehagias}}\ and\ \bibinfo {author} {\bibfnamefont {A.}~\bibnamefont
  {Riotto}},\ }\href {https://doi.org/10.1103/j5fw-w3xc} {\bibfield  {journal}
  {\bibinfo  {journal} {Phys. Rev. D}\ }\textbf {\bibinfo {volume} {112}},\
  \bibinfo {pages} {024023} (\bibinfo {year} {2025})},\ \Eprint
  {https://arxiv.org/abs/2504.06224} {arXiv:2504.06224 [gr-qc]} \BibitemShut
  {NoStop}%
\bibitem [{\citenamefont {He}\ \emph {et~al.}(2025)\citenamefont {He},
  \citenamefont {Du}, \citenamefont {Jiao}, \citenamefont {Shao}, \citenamefont
  {Shi}, \citenamefont {Tian},\ and\ \citenamefont {Zhang}}]{He:2025}%
  \BibitemOpen
  \bibfield  {author} {\bibinfo {author} {\bibfnamefont {Z.-T.}\ \bibnamefont
  {He}}, \bibinfo {author} {\bibfnamefont {J.}~\bibnamefont {Du}}, \bibinfo
  {author} {\bibfnamefont {J.}~\bibnamefont {Jiao}}, \bibinfo {author}
  {\bibfnamefont {C.}~\bibnamefont {Shao}}, \bibinfo {author} {\bibfnamefont
  {J.}~\bibnamefont {Shi}}, \bibinfo {author} {\bibfnamefont {Y.}~\bibnamefont
  {Tian}},\ and\ \bibinfo {author} {\bibfnamefont {H.}~\bibnamefont {Zhang}},\
  }\href {https://doi.org/10.1103/xrs3-82cb} {\bibfield  {journal} {\bibinfo
  {journal} {Phys. Rev. D}\ }\textbf {\bibinfo {volume} {112}},\ \bibinfo
  {pages} {104008} (\bibinfo {year} {2025})},\ \Eprint
  {https://arxiv.org/abs/2508.20499} {arXiv:2508.20499 [gr-qc]} \BibitemShut
  {NoStop}%
\bibitem [{\citenamefont {Shao}\ \emph {et~al.}(2026)\citenamefont {Shao},
  \citenamefont {He}, \citenamefont {Jiao}, \citenamefont {Lai}, \citenamefont
  {Shi}, \citenamefont {Tian}, \citenamefont {Yuan},\ and\ \citenamefont
  {Zhang}}]{Shao:2026}%
  \BibitemOpen
  \bibfield  {author} {\bibinfo {author} {\bibfnamefont {C.}~\bibnamefont
  {Shao}}, \bibinfo {author} {\bibfnamefont {Z.-T.}\ \bibnamefont {He}},
  \bibinfo {author} {\bibfnamefont {J.}~\bibnamefont {Jiao}}, \bibinfo {author}
  {\bibfnamefont {J.}~\bibnamefont {Lai}}, \bibinfo {author} {\bibfnamefont
  {J.-X.}\ \bibnamefont {Shi}}, \bibinfo {author} {\bibfnamefont
  {Y.}~\bibnamefont {Tian}}, \bibinfo {author} {\bibfnamefont {D.}~\bibnamefont
  {Yuan}},\ and\ \bibinfo {author} {\bibfnamefont {H.}~\bibnamefont {Zhang}},\
  }\href@noop {} {\  (\bibinfo {year} {2026})},\ \Eprint
  {https://arxiv.org/abs/2601.16016} {arXiv:2601.16016 [gr-qc]} \BibitemShut
  {NoStop}%
\bibitem [{\citenamefont {Redondo-Yuste}\ \emph {et~al.}(2024)\citenamefont
  {Redondo-Yuste}, \citenamefont {Carullo}, \citenamefont {Ripley},
  \citenamefont {Berti},\ and\ \citenamefont {Cardoso}}]{Yuste:2024}%
  \BibitemOpen
  \bibfield  {author} {\bibinfo {author} {\bibfnamefont {J.}~\bibnamefont
  {Redondo-Yuste}}, \bibinfo {author} {\bibfnamefont {G.}~\bibnamefont
  {Carullo}}, \bibinfo {author} {\bibfnamefont {J.~L.}\ \bibnamefont {Ripley}},
  \bibinfo {author} {\bibfnamefont {E.}~\bibnamefont {Berti}},\ and\ \bibinfo
  {author} {\bibfnamefont {V.}~\bibnamefont {Cardoso}},\ }\href
  {https://doi.org/10.1103/PhysRevD.109.L101503} {\bibfield  {journal}
  {\bibinfo  {journal} {Phys. Rev. D}\ }\textbf {\bibinfo {volume} {109}},\
  \bibinfo {pages} {L101503} (\bibinfo {year} {2024})}\BibitemShut {NoStop}%
\bibitem [{\citenamefont {Cheung}\ \emph
  {et~al.}(2024{\natexlab{a}})\citenamefont {Cheung}, \citenamefont {Berti},
  \citenamefont {Baibhav},\ and\ \citenamefont {Cotesta}}]{Cheung:2024}%
  \BibitemOpen
  \bibfield  {author} {\bibinfo {author} {\bibfnamefont {M.~H.-Y.}\
  \bibnamefont {Cheung}}, \bibinfo {author} {\bibfnamefont {E.}~\bibnamefont
  {Berti}}, \bibinfo {author} {\bibfnamefont {V.}~\bibnamefont {Baibhav}},\
  and\ \bibinfo {author} {\bibfnamefont {R.}~\bibnamefont {Cotesta}},\ }\href
  {https://doi.org/10.1103/PhysRevD.109.044069} {\bibfield  {journal} {\bibinfo
   {journal} {Phys. Rev. D}\ }\textbf {\bibinfo {volume} {109}},\ \bibinfo
  {pages} {044069} (\bibinfo {year} {2024}{\natexlab{a}})}\BibitemShut
  {NoStop}%
\bibitem [{\citenamefont {Ma}\ and\ \citenamefont {Yang}(2024)}]{Ma:2024}%
  \BibitemOpen
  \bibfield  {author} {\bibinfo {author} {\bibfnamefont {S.}~\bibnamefont
  {Ma}}\ and\ \bibinfo {author} {\bibfnamefont {H.}~\bibnamefont {Yang}},\
  }\href {https://doi.org/10.1103/PhysRevD.109.104070} {\bibfield  {journal}
  {\bibinfo  {journal} {Phys. Rev. D}\ }\textbf {\bibinfo {volume} {109}},\
  \bibinfo {pages} {104070} (\bibinfo {year} {2024})}\BibitemShut {NoStop}%
\bibitem [{\citenamefont {Zhu}\ \emph {et~al.}(2024)\citenamefont {Zhu},
  \citenamefont {Ripley}, \citenamefont {Pretorius}, \citenamefont {Ma},
  \citenamefont {Mitman}, \citenamefont {Owen}, \citenamefont {Boyle},
  \citenamefont {Chen}, \citenamefont {Deppe}, \citenamefont {Kidder},
  \citenamefont {Moxon}, \citenamefont {Nelli}, \citenamefont {Pfeiffer},
  \citenamefont {Scheel}, \citenamefont {Throwe},\ and\ \citenamefont
  {Vu}}]{Zhu:2024}%
  \BibitemOpen
  \bibfield  {author} {\bibinfo {author} {\bibfnamefont {H.}~\bibnamefont
  {Zhu}}, \bibinfo {author} {\bibfnamefont {J.~L.}\ \bibnamefont {Ripley}},
  \bibinfo {author} {\bibfnamefont {F.}~\bibnamefont {Pretorius}}, \bibinfo
  {author} {\bibfnamefont {S.}~\bibnamefont {Ma}}, \bibinfo {author}
  {\bibfnamefont {K.}~\bibnamefont {Mitman}}, \bibinfo {author} {\bibfnamefont
  {R.}~\bibnamefont {Owen}}, \bibinfo {author} {\bibfnamefont {M.}~\bibnamefont
  {Boyle}}, \bibinfo {author} {\bibfnamefont {Y.}~\bibnamefont {Chen}},
  \bibinfo {author} {\bibfnamefont {N.}~\bibnamefont {Deppe}}, \bibinfo
  {author} {\bibfnamefont {L.~E.}\ \bibnamefont {Kidder}}, \bibinfo {author}
  {\bibfnamefont {J.}~\bibnamefont {Moxon}}, \bibinfo {author} {\bibfnamefont
  {K.~C.}\ \bibnamefont {Nelli}}, \bibinfo {author} {\bibfnamefont {H.~P.}\
  \bibnamefont {Pfeiffer}}, \bibinfo {author} {\bibfnamefont {M.~A.}\
  \bibnamefont {Scheel}}, \bibinfo {author} {\bibfnamefont {W.}~\bibnamefont
  {Throwe}},\ and\ \bibinfo {author} {\bibfnamefont {N.~L.}\ \bibnamefont
  {Vu}},\ }\href {https://doi.org/10.1103/PhysRevD.109.104050} {\bibfield
  {journal} {\bibinfo  {journal} {Phys. Rev. D}\ }\textbf {\bibinfo {volume}
  {109}},\ \bibinfo {pages} {104050} (\bibinfo {year} {2024})}\BibitemShut
  {NoStop}%
\bibitem [{\citenamefont {Bucciotti}\ \emph {et~al.}(2024)\citenamefont
  {Bucciotti}, \citenamefont {Juliano}, \citenamefont {Kuntz},\ and\
  \citenamefont {Trincherini}}]{Buccio:2024}%
  \BibitemOpen
  \bibfield  {author} {\bibinfo {author} {\bibfnamefont {B.}~\bibnamefont
  {Bucciotti}}, \bibinfo {author} {\bibfnamefont {L.}~\bibnamefont {Juliano}},
  \bibinfo {author} {\bibfnamefont {A.}~\bibnamefont {Kuntz}},\ and\ \bibinfo
  {author} {\bibfnamefont {E.}~\bibnamefont {Trincherini}},\ }\href
  {https://doi.org/10.1103/PhysRevD.110.104048} {\bibfield  {journal} {\bibinfo
   {journal} {Phys. Rev. D}\ }\textbf {\bibinfo {volume} {110}},\ \bibinfo
  {pages} {104048} (\bibinfo {year} {2024})}\BibitemShut {NoStop}%
\bibitem [{\citenamefont {Bourg}\ \emph {et~al.}(2025)\citenamefont {Bourg},
  \citenamefont {Macedo}, \citenamefont {Spiers}, \citenamefont {Leather},
  \citenamefont {Bonga},\ and\ \citenamefont {Pound}}]{Bourg:2025}%
  \BibitemOpen
  \bibfield  {author} {\bibinfo {author} {\bibfnamefont {P.}~\bibnamefont
  {Bourg}}, \bibinfo {author} {\bibfnamefont {R.~P.}\ \bibnamefont {Macedo}},
  \bibinfo {author} {\bibfnamefont {A.}~\bibnamefont {Spiers}}, \bibinfo
  {author} {\bibfnamefont {B.}~\bibnamefont {Leather}}, \bibinfo {author}
  {\bibfnamefont {B.}~\bibnamefont {Bonga}},\ and\ \bibinfo {author}
  {\bibfnamefont {A.}~\bibnamefont {Pound}},\ }\href
  {https://doi.org/10.1103/PhysRevLett.134.061401} {\bibfield  {journal}
  {\bibinfo  {journal} {Phys. Rev. Lett.}\ }\textbf {\bibinfo {volume} {134}},\
  \bibinfo {pages} {061401} (\bibinfo {year} {2025})}\BibitemShut {NoStop}%
\bibitem [{\citenamefont {Khera}\ \emph {et~al.}(2025)\citenamefont {Khera},
  \citenamefont {Ma},\ and\ \citenamefont {Yang}}]{Khera:2025}%
  \BibitemOpen
  \bibfield  {author} {\bibinfo {author} {\bibfnamefont {N.}~\bibnamefont
  {Khera}}, \bibinfo {author} {\bibfnamefont {S.}~\bibnamefont {Ma}},\ and\
  \bibinfo {author} {\bibfnamefont {H.}~\bibnamefont {Yang}},\ }\href
  {https://doi.org/10.1103/PhysRevLett.134.211404} {\bibfield  {journal}
  {\bibinfo  {journal} {Phys. Rev. Lett.}\ }\textbf {\bibinfo {volume} {134}},\
  \bibinfo {pages} {211404} (\bibinfo {year} {2025})}\BibitemShut {NoStop}%
\bibitem [{\citenamefont {Yang}\ \emph {et~al.}(2025)\citenamefont {Yang},
  \citenamefont {Shi},\ and\ \citenamefont {Hu}}]{Yang:2025}%
  \BibitemOpen
  \bibfield  {author} {\bibinfo {author} {\bibfnamefont {Y.}~\bibnamefont
  {Yang}}, \bibinfo {author} {\bibfnamefont {C.}~\bibnamefont {Shi}},\ and\
  \bibinfo {author} {\bibfnamefont {Y.-M.}\ \bibnamefont {Hu}},\ }\href
  {https://arxiv.org/abs/2510.16903} {\bibinfo {title} {Contribution from
  nonlinear quasi-normal modes in gw250114}} (\bibinfo {year} {2025}),\ \Eprint
  {https://arxiv.org/abs/2510.16903} {arXiv:2510.16903 [gr-qc]} \BibitemShut
  {NoStop}%
\bibitem [{\citenamefont {Wang}\ \emph {et~al.}(2026)\citenamefont {Wang},
  \citenamefont {Ma}, \citenamefont {Khera},\ and\ \citenamefont
  {Yang}}]{Wang:2026rev}%
  \BibitemOpen
  \bibfield  {author} {\bibinfo {author} {\bibfnamefont {Y.-F.}\ \bibnamefont
  {Wang}}, \bibinfo {author} {\bibfnamefont {S.}~\bibnamefont {Ma}}, \bibinfo
  {author} {\bibfnamefont {N.}~\bibnamefont {Khera}},\ and\ \bibinfo {author}
  {\bibfnamefont {H.}~\bibnamefont {Yang}},\ }\href@noop {} {\  (\bibinfo
  {year} {2026})},\ \Eprint {https://arxiv.org/abs/2601.05734}
  {arXiv:2601.05734 [gr-qc]} \BibitemShut {NoStop}%
\bibitem [{\citenamefont {Strominger}\ and\ \citenamefont
  {Zhiboedov}(2016)}]{StromingerZhiboedov2016}%
  \BibitemOpen
  \bibfield  {author} {\bibinfo {author} {\bibfnamefont {A.}~\bibnamefont
  {Strominger}}\ and\ \bibinfo {author} {\bibfnamefont {A.}~\bibnamefont
  {Zhiboedov}},\ }\href {https://doi.org/10.1007/JHEP01(2016)086} {\bibfield
  {journal} {\bibinfo  {journal} {Journal of High Energy Physics}\ }\textbf
  {\bibinfo {volume} {01}},\ \bibinfo {pages} {086} (\bibinfo {year} {2016})},\
  \Eprint {https://arxiv.org/abs/1411.5745} {arXiv:1411.5745 [hep-th]}
  \BibitemShut {NoStop}%
\bibitem [{\citenamefont {Strominger}(2017)}]{Strominger2017LecturesIR}%
  \BibitemOpen
  \bibfield  {author} {\bibinfo {author} {\bibfnamefont {A.}~\bibnamefont
  {Strominger}},\ }\href@noop {} {\bibfield  {journal} {\bibinfo  {journal}
  {arXiv preprint}\ } (\bibinfo {year} {2017})},\ \Eprint
  {https://arxiv.org/abs/1703.05448} {arXiv:1703.05448 [hep-th]} \BibitemShut
  {NoStop}%
\bibitem [{\citenamefont {Cheung}\ \emph
  {et~al.}(2024{\natexlab{b}})\citenamefont {Cheung}, \citenamefont {Lasky},\
  and\ \citenamefont {Thrane}}]{Cheung:2024memory}%
  \BibitemOpen
  \bibfield  {author} {\bibinfo {author} {\bibfnamefont {S.~Y.}\ \bibnamefont
  {Cheung}}, \bibinfo {author} {\bibfnamefont {P.~D.}\ \bibnamefont {Lasky}},\
  and\ \bibinfo {author} {\bibfnamefont {E.}~\bibnamefont {Thrane}},\ }\href
  {https://doi.org/10.1088/1361-6382/ad3ffe} {\bibfield  {journal} {\bibinfo
  {journal} {Class. Quant. Grav.}\ }\textbf {\bibinfo {volume} {41}},\ \bibinfo
  {pages} {115010} (\bibinfo {year} {2024}{\natexlab{b}})},\ \Eprint
  {https://arxiv.org/abs/2404.11919} {arXiv:2404.11919 [gr-qc]} \BibitemShut
  {NoStop}%
\bibitem [{\citenamefont {H\"ubner}\ \emph {et~al.}(2020)\citenamefont
  {H\"ubner}, \citenamefont {Talbot}, \citenamefont {Lasky},\ and\
  \citenamefont {Thrane}}]{PhysRevD.101.023011}%
  \BibitemOpen
  \bibfield  {author} {\bibinfo {author} {\bibfnamefont {M.}~\bibnamefont
  {H\"ubner}}, \bibinfo {author} {\bibfnamefont {C.}~\bibnamefont {Talbot}},
  \bibinfo {author} {\bibfnamefont {P.~D.}\ \bibnamefont {Lasky}},\ and\
  \bibinfo {author} {\bibfnamefont {E.}~\bibnamefont {Thrane}},\ }\href
  {https://doi.org/10.1103/PhysRevD.101.023011} {\bibfield  {journal} {\bibinfo
   {journal} {Phys. Rev. D}\ }\textbf {\bibinfo {volume} {101}},\ \bibinfo
  {pages} {023011} (\bibinfo {year} {2020})}\BibitemShut {NoStop}%
\bibitem [{\citenamefont {Lasky}\ \emph {et~al.}(2016)\citenamefont {Lasky},
  \citenamefont {Thrane}, \citenamefont {Levin}, \citenamefont {Blackman},\
  and\ \citenamefont {Chen}}]{Lasky:2016}%
  \BibitemOpen
  \bibfield  {author} {\bibinfo {author} {\bibfnamefont {P.~D.}\ \bibnamefont
  {Lasky}}, \bibinfo {author} {\bibfnamefont {E.}~\bibnamefont {Thrane}},
  \bibinfo {author} {\bibfnamefont {Y.}~\bibnamefont {Levin}}, \bibinfo
  {author} {\bibfnamefont {J.}~\bibnamefont {Blackman}},\ and\ \bibinfo
  {author} {\bibfnamefont {Y.}~\bibnamefont {Chen}},\ }\href
  {https://doi.org/10.1103/PhysRevLett.117.061102} {\bibfield  {journal}
  {\bibinfo  {journal} {Phys. Rev. Lett.}\ }\textbf {\bibinfo {volume} {117}},\
  \bibinfo {pages} {061102} (\bibinfo {year} {2016})}\BibitemShut {NoStop}%
\bibitem [{\citenamefont {Zhao}\ \emph {et~al.}(2021)\citenamefont {Zhao},
  \citenamefont {Liu}, \citenamefont {Cao},\ and\ \citenamefont
  {He}}]{Zhao:2021}%
  \BibitemOpen
  \bibfield  {author} {\bibinfo {author} {\bibfnamefont {Z.-C.}\ \bibnamefont
  {Zhao}}, \bibinfo {author} {\bibfnamefont {X.}~\bibnamefont {Liu}}, \bibinfo
  {author} {\bibfnamefont {Z.}~\bibnamefont {Cao}},\ and\ \bibinfo {author}
  {\bibfnamefont {X.}~\bibnamefont {He}},\ }\href
  {https://doi.org/10.1103/PhysRevD.104.064056} {\bibfield  {journal} {\bibinfo
   {journal} {Phys. Rev. D}\ }\textbf {\bibinfo {volume} {104}},\ \bibinfo
  {pages} {064056} (\bibinfo {year} {2021})}\BibitemShut {NoStop}%
\bibitem [{\citenamefont {Ebersold}\ and\ \citenamefont
  {Tiwari}(2020)}]{Ebersold:2020}%
  \BibitemOpen
  \bibfield  {author} {\bibinfo {author} {\bibfnamefont {M.}~\bibnamefont
  {Ebersold}}\ and\ \bibinfo {author} {\bibfnamefont {S.}~\bibnamefont
  {Tiwari}},\ }\href {https://doi.org/10.1103/PhysRevD.101.104041} {\bibfield
  {journal} {\bibinfo  {journal} {Phys. Rev. D}\ }\textbf {\bibinfo {volume}
  {101}},\ \bibinfo {pages} {104041} (\bibinfo {year} {2020})},\ \Eprint
  {https://arxiv.org/abs/2005.03306} {arXiv:2005.03306 [gr-qc]} \BibitemShut
  {NoStop}%
\bibitem [{\citenamefont {Babak}\ \emph {et~al.}(2024)\citenamefont {Babak},
  \citenamefont {Falxa}, \citenamefont {Franciolini},\ and\ \citenamefont
  {Pieroni}}]{BabakPTA:2024}%
  \BibitemOpen
  \bibfield  {author} {\bibinfo {author} {\bibfnamefont {S.}~\bibnamefont
  {Babak}}, \bibinfo {author} {\bibfnamefont {M.}~\bibnamefont {Falxa}},
  \bibinfo {author} {\bibfnamefont {G.}~\bibnamefont {Franciolini}},\ and\
  \bibinfo {author} {\bibfnamefont {M.}~\bibnamefont {Pieroni}},\ }\href
  {https://doi.org/10.1103/PhysRevD.110.063022} {\bibfield  {journal} {\bibinfo
   {journal} {Phys. Rev. D}\ }\textbf {\bibinfo {volume} {110}},\ \bibinfo
  {pages} {063022} (\bibinfo {year} {2024})},\ \Eprint
  {https://arxiv.org/abs/2404.02864} {arXiv:2404.02864 [astro-ph.CO]}
  \BibitemShut {NoStop}%
\bibitem [{\citenamefont {Inchausp{\'e}}\ \emph {et~al.}(2025)\citenamefont
  {Inchausp{\'e}}, \citenamefont {Gasparotto}, \citenamefont {Blas},
  \citenamefont {Heisenberg}, \citenamefont {Zosso},\ and\ \citenamefont
  {Tiwari}}]{Incha:2024}%
  \BibitemOpen
  \bibfield  {author} {\bibinfo {author} {\bibfnamefont {H.}~\bibnamefont
  {Inchausp{\'e}}}, \bibinfo {author} {\bibfnamefont {S.}~\bibnamefont
  {Gasparotto}}, \bibinfo {author} {\bibfnamefont {D.}~\bibnamefont {Blas}},
  \bibinfo {author} {\bibfnamefont {L.}~\bibnamefont {Heisenberg}}, \bibinfo
  {author} {\bibfnamefont {J.}~\bibnamefont {Zosso}},\ and\ \bibinfo {author}
  {\bibfnamefont {S.}~\bibnamefont {Tiwari}},\ }\href
  {https://doi.org/10.1103/PhysRevD.111.044044} {\bibfield  {journal} {\bibinfo
   {journal} {Phys. Rev. D}\ }\textbf {\bibinfo {volume} {111}},\ \bibinfo
  {pages} {044044} (\bibinfo {year} {2025})},\ \Eprint
  {https://arxiv.org/abs/2406.09228} {arXiv:2406.09228 [gr-qc]} \BibitemShut
  {NoStop}%
\bibitem [{\citenamefont {Ghosh}\ \emph {et~al.}(2023)\citenamefont {Ghosh},
  \citenamefont {Weaver}, \citenamefont {Sanjuan}, \citenamefont {Fulda},\ and\
  \citenamefont {Mueller}}]{Ghosh:2023}%
  \BibitemOpen
  \bibfield  {author} {\bibinfo {author} {\bibfnamefont {S.}~\bibnamefont
  {Ghosh}}, \bibinfo {author} {\bibfnamefont {A.}~\bibnamefont {Weaver}},
  \bibinfo {author} {\bibfnamefont {J.}~\bibnamefont {Sanjuan}}, \bibinfo
  {author} {\bibfnamefont {P.}~\bibnamefont {Fulda}},\ and\ \bibinfo {author}
  {\bibfnamefont {G.}~\bibnamefont {Mueller}},\ }\href
  {https://doi.org/10.1103/PhysRevD.107.084051} {\bibfield  {journal} {\bibinfo
   {journal} {Phys. Rev. D}\ }\textbf {\bibinfo {volume} {107}},\ \bibinfo
  {pages} {084051} (\bibinfo {year} {2023})},\ \Eprint
  {https://arxiv.org/abs/2302.04396} {arXiv:2302.04396 [gr-qc]} \BibitemShut
  {NoStop}%
\bibitem [{\citenamefont {Lagos}\ and\ \citenamefont {Hui}(2023)}]{qqnmfreq1}%
  \BibitemOpen
  \bibfield  {author} {\bibinfo {author} {\bibfnamefont {M.}~\bibnamefont
  {Lagos}}\ and\ \bibinfo {author} {\bibfnamefont {L.}~\bibnamefont {Hui}},\
  }\href {https://doi.org/10.1103/PhysRevD.107.044040} {\bibfield  {journal}
  {\bibinfo  {journal} {Phys. Rev. D}\ }\textbf {\bibinfo {volume} {107}},\
  \bibinfo {pages} {044040} (\bibinfo {year} {2023})}\BibitemShut {NoStop}%
\bibitem [{\citenamefont {Pazos}\ \emph {et~al.}(2010)\citenamefont {Pazos},
  \citenamefont {Brizuela}, \citenamefont {Mart\'{\i}n-Garc\'{\i}a},\ and\
  \citenamefont {Tiglio}}]{qqnmfreq2}%
  \BibitemOpen
  \bibfield  {author} {\bibinfo {author} {\bibfnamefont {E.}~\bibnamefont
  {Pazos}}, \bibinfo {author} {\bibfnamefont {D.}~\bibnamefont {Brizuela}},
  \bibinfo {author} {\bibfnamefont {J.~M.}\ \bibnamefont
  {Mart\'{\i}n-Garc\'{\i}a}},\ and\ \bibinfo {author} {\bibfnamefont
  {M.}~\bibnamefont {Tiglio}},\ }\href
  {https://doi.org/10.1103/PhysRevD.82.104028} {\bibfield  {journal} {\bibinfo
  {journal} {Phys. Rev. D}\ }\textbf {\bibinfo {volume} {82}},\ \bibinfo
  {pages} {104028} (\bibinfo {year} {2010})}\BibitemShut {NoStop}%
\bibitem [{\citenamefont {London}\ \emph
  {et~al.}(2014{\natexlab{a}})\citenamefont {London}, \citenamefont
  {Shoemaker},\ and\ \citenamefont {Healy}}]{qqnmfreq3}%
  \BibitemOpen
  \bibfield  {author} {\bibinfo {author} {\bibfnamefont {L.}~\bibnamefont
  {London}}, \bibinfo {author} {\bibfnamefont {D.}~\bibnamefont {Shoemaker}},\
  and\ \bibinfo {author} {\bibfnamefont {J.}~\bibnamefont {Healy}},\ }\href
  {https://doi.org/10.1103/PhysRevD.90.124032} {\bibfield  {journal} {\bibinfo
  {journal} {Phys. Rev. D}\ }\textbf {\bibinfo {volume} {90}},\ \bibinfo
  {pages} {124032} (\bibinfo {year} {2014}{\natexlab{a}})}\BibitemShut
  {NoStop}%
\bibitem [{\citenamefont {London}\ \emph
  {et~al.}(2014{\natexlab{b}})\citenamefont {London}, \citenamefont
  {Shoemaker},\ and\ \citenamefont {Healy}}]{London:2014}%
  \BibitemOpen
  \bibfield  {author} {\bibinfo {author} {\bibfnamefont {L.}~\bibnamefont
  {London}}, \bibinfo {author} {\bibfnamefont {D.}~\bibnamefont {Shoemaker}},\
  and\ \bibinfo {author} {\bibfnamefont {J.}~\bibnamefont {Healy}},\ }\href
  {https://doi.org/10.1103/PhysRevD.90.124032} {\bibfield  {journal} {\bibinfo
  {journal} {Phys. Rev. D}\ }\textbf {\bibinfo {volume} {90}},\ \bibinfo
  {pages} {124032} (\bibinfo {year} {2014}{\natexlab{b}})},\ \bibinfo {note}
  {[Erratum: Phys.Rev.D 94, 069902 (2016)]},\ \Eprint
  {https://arxiv.org/abs/1404.3197} {arXiv:1404.3197 [gr-qc]} \BibitemShut
  {NoStop}%
\bibitem [{\citenamefont
  {Favata}(2009{\natexlab{a}})}]{Favata2009NonlinearBBH}%
  \BibitemOpen
  \bibfield  {author} {\bibinfo {author} {\bibfnamefont {M.}~\bibnamefont
  {Favata}},\ }\href {https://doi.org/10.1088/0004-637X/696/2/L159} {\bibfield
  {journal} {\bibinfo  {journal} {Astrophysical Journal Letters}\ }\textbf
  {\bibinfo {volume} {696}},\ \bibinfo {pages} {L159} (\bibinfo {year}
  {2009}{\natexlab{a}})},\ \Eprint {https://arxiv.org/abs/0902.3660}
  {arXiv:0902.3660 [astro-ph]} \BibitemShut {NoStop}%
\bibitem [{\citenamefont {Favata}(2009{\natexlab{b}})}]{Favata2009Revisited}%
  \BibitemOpen
  \bibfield  {author} {\bibinfo {author} {\bibfnamefont {M.}~\bibnamefont
  {Favata}},\ }in\ \href {https://doi.org/10.1088/1742-6596/154/1/012043}
  {\emph {\bibinfo {booktitle} {Journal of Physics: Conference Series}}},\
  Vol.\ \bibinfo {volume} {154}\ (\bibinfo {year} {2009})\ p.\ \bibinfo {pages}
  {012043},\ \Eprint {https://arxiv.org/abs/0811.3451} {arXiv:0811.3451
  [astro-ph]} \BibitemShut {NoStop}%
\bibitem [{\citenamefont {Boyle}\ \emph {et~al.}(2019)\citenamefont {Boyle}
  \emph {et~al.}}]{sxs1}%
  \BibitemOpen
  \bibfield  {author} {\bibinfo {author} {\bibfnamefont {M.}~\bibnamefont
  {Boyle}} \emph {et~al.},\ }\href {https://doi.org/10.1088/1361-6382/ab34e2}
  {\bibfield  {journal} {\bibinfo  {journal} {Class. Quant. Grav.}\ }\textbf
  {\bibinfo {volume} {36}},\ \bibinfo {pages} {195006} (\bibinfo {year}
  {2019})},\ \Eprint {https://arxiv.org/abs/1904.04831} {arXiv:1904.04831
  [gr-qc]} \BibitemShut {NoStop}%
\bibitem [{\citenamefont {Scheel}\ \emph {et~al.}(2025)\citenamefont {Scheel}
  \emph {et~al.}}]{sxs2}%
  \BibitemOpen
  \bibfield  {author} {\bibinfo {author} {\bibfnamefont {M.~A.}\ \bibnamefont
  {Scheel}} \emph {et~al.},\ }\href {https://doi.org/10.1088/1361-6382/adfd34}
  {\bibfield  {journal} {\bibinfo  {journal} {Class. Quant. Grav.}\ }\textbf
  {\bibinfo {volume} {42}},\ \bibinfo {pages} {195017} (\bibinfo {year}
  {2025})},\ \Eprint {https://arxiv.org/abs/2505.13378} {arXiv:2505.13378
  [gr-qc]} \BibitemShut {NoStop}%
\bibitem [{\citenamefont {{SXS Collaboration}}({\natexlab{a}})}]{sxs3}%
  \BibitemOpen
  \bibfield  {author} {\bibinfo {author} {\bibnamefont {{SXS Collaboration}}},\
  }\href@noop {} {\bibinfo {title} {{SpEC: Spectral Einstein Code}}},\ \bibinfo
  {howpublished} {\url{https://www.black-holes.org/code/SpEC.html}}
  ({\natexlab{a}}),\ \bibinfo {note} {accessed: 2025-12-20}\BibitemShut
  {NoStop}%
\bibitem [{\citenamefont {{SXS Collaboration}}({\natexlab{b}})}]{sxs4}%
  \BibitemOpen
  \bibfield  {author} {\bibinfo {author} {\bibnamefont {{SXS Collaboration}}},\
  }\href@noop {} {\bibinfo {title} {{SXS Gravitational Waveform Database}}},\
  \bibinfo {howpublished} {\url{https://www.black-holes.org/waveforms}}
  ({\natexlab{b}}),\ \bibinfo {note} {accessed: 2025-12-20}\BibitemShut
  {NoStop}%
\bibitem [{\citenamefont {Moxon}\ \emph {et~al.}(2020)\citenamefont {Moxon},
  \citenamefont {Scheel},\ and\ \citenamefont {Teukolsky}}]{cce1}%
  \BibitemOpen
  \bibfield  {author} {\bibinfo {author} {\bibfnamefont {J.}~\bibnamefont
  {Moxon}}, \bibinfo {author} {\bibfnamefont {M.~A.}\ \bibnamefont {Scheel}},\
  and\ \bibinfo {author} {\bibfnamefont {S.~A.}\ \bibnamefont {Teukolsky}},\
  }\href {https://doi.org/10.1103/PhysRevD.102.044052} {\bibfield  {journal}
  {\bibinfo  {journal} {Phys. Rev. D}\ }\textbf {\bibinfo {volume} {102}},\
  \bibinfo {pages} {044052} (\bibinfo {year} {2020})}\BibitemShut {NoStop}%
\bibitem [{\citenamefont {Moxon}\ \emph {et~al.}(2023)\citenamefont {Moxon},
  \citenamefont {Scheel}, \citenamefont {Teukolsky}, \citenamefont {Deppe},
  \citenamefont {Fischer}, \citenamefont {H{\'e}bert}, \citenamefont {Kidder},\
  and\ \citenamefont {Throwe}}]{cce2}%
  \BibitemOpen
  \bibfield  {author} {\bibinfo {author} {\bibfnamefont {J.}~\bibnamefont
  {Moxon}}, \bibinfo {author} {\bibfnamefont {M.~A.}\ \bibnamefont {Scheel}},
  \bibinfo {author} {\bibfnamefont {S.~A.}\ \bibnamefont {Teukolsky}}, \bibinfo
  {author} {\bibfnamefont {N.}~\bibnamefont {Deppe}}, \bibinfo {author}
  {\bibfnamefont {N.}~\bibnamefont {Fischer}}, \bibinfo {author} {\bibfnamefont
  {F.}~\bibnamefont {H{\'e}bert}}, \bibinfo {author} {\bibfnamefont {L.~E.}\
  \bibnamefont {Kidder}},\ and\ \bibinfo {author} {\bibfnamefont
  {W.}~\bibnamefont {Throwe}},\ }\href
  {https://doi.org/10.1103/PhysRevD.107.064013} {\bibfield  {journal} {\bibinfo
   {journal} {Phys. Rev. D}\ }\textbf {\bibinfo {volume} {107}},\ \bibinfo
  {pages} {064013} (\bibinfo {year} {2023})},\ \Eprint
  {https://arxiv.org/abs/2110.08635} {arXiv:2110.08635 [gr-qc]} \BibitemShut
  {NoStop}%
\bibitem [{\citenamefont {Mitman}\ \emph {et~al.}(2021)\citenamefont {Mitman},
  \citenamefont {Khera}, \citenamefont {Iozzo}, \citenamefont {Stein},
  \citenamefont {Boyle}, \citenamefont {Deppe}, \citenamefont {Kidder},
  \citenamefont {Moxon}, \citenamefont {Pfeiffer}, \citenamefont {Scheel},
  \citenamefont {Teukolsky},\ and\ \citenamefont {Throwe}}]{cce3}%
  \BibitemOpen
  \bibfield  {author} {\bibinfo {author} {\bibfnamefont {K.}~\bibnamefont
  {Mitman}}, \bibinfo {author} {\bibfnamefont {N.}~\bibnamefont {Khera}},
  \bibinfo {author} {\bibfnamefont {D.~A.~B.}\ \bibnamefont {Iozzo}}, \bibinfo
  {author} {\bibfnamefont {L.~C.}\ \bibnamefont {Stein}}, \bibinfo {author}
  {\bibfnamefont {M.}~\bibnamefont {Boyle}}, \bibinfo {author} {\bibfnamefont
  {N.}~\bibnamefont {Deppe}}, \bibinfo {author} {\bibfnamefont {L.~E.}\
  \bibnamefont {Kidder}}, \bibinfo {author} {\bibfnamefont {J.}~\bibnamefont
  {Moxon}}, \bibinfo {author} {\bibfnamefont {H.~P.}\ \bibnamefont {Pfeiffer}},
  \bibinfo {author} {\bibfnamefont {M.~A.}\ \bibnamefont {Scheel}}, \bibinfo
  {author} {\bibfnamefont {S.~A.}\ \bibnamefont {Teukolsky}},\ and\ \bibinfo
  {author} {\bibfnamefont {W.}~\bibnamefont {Throwe}},\ }\href
  {https://doi.org/10.1103/PhysRevD.104.024051} {\bibfield  {journal} {\bibinfo
   {journal} {Phys. Rev. D}\ }\textbf {\bibinfo {volume} {104}},\ \bibinfo
  {pages} {024051} (\bibinfo {year} {2021})}\BibitemShut {NoStop}%
\bibitem [{\citenamefont {Mitman}\ \emph {et~al.}(2022)\citenamefont {Mitman},
  \citenamefont {Stein}, \citenamefont {Boyle}, \citenamefont {Deppe},
  \citenamefont {H\'ebert}, \citenamefont {Kidder}, \citenamefont {Moxon},
  \citenamefont {Scheel}, \citenamefont {Teukolsky}, \citenamefont {Throwe},\
  and\ \citenamefont {Vu}}]{cce4}%
  \BibitemOpen
  \bibfield  {author} {\bibinfo {author} {\bibfnamefont {K.}~\bibnamefont
  {Mitman}}, \bibinfo {author} {\bibfnamefont {L.~C.}\ \bibnamefont {Stein}},
  \bibinfo {author} {\bibfnamefont {M.}~\bibnamefont {Boyle}}, \bibinfo
  {author} {\bibfnamefont {N.}~\bibnamefont {Deppe}}, \bibinfo {author}
  {\bibfnamefont {F.~m.~c.}\ \bibnamefont {H\'ebert}}, \bibinfo {author}
  {\bibfnamefont {L.~E.}\ \bibnamefont {Kidder}}, \bibinfo {author}
  {\bibfnamefont {J.}~\bibnamefont {Moxon}}, \bibinfo {author} {\bibfnamefont
  {M.~A.}\ \bibnamefont {Scheel}}, \bibinfo {author} {\bibfnamefont {S.~A.}\
  \bibnamefont {Teukolsky}}, \bibinfo {author} {\bibfnamefont {W.}~\bibnamefont
  {Throwe}},\ and\ \bibinfo {author} {\bibfnamefont {N.~L.}\ \bibnamefont
  {Vu}},\ }\href {https://doi.org/10.1103/PhysRevD.106.084029} {\bibfield
  {journal} {\bibinfo  {journal} {Phys. Rev. D}\ }\textbf {\bibinfo {volume}
  {106}},\ \bibinfo {pages} {084029} (\bibinfo {year} {2022})}\BibitemShut
  {NoStop}%
\bibitem [{\citenamefont {Boyle}\ \emph {et~al.}(2020)\citenamefont {Boyle},
  \citenamefont {Iozzo},\ and\ \citenamefont {Stein}}]{scri}%
  \BibitemOpen
  \bibfield  {author} {\bibinfo {author} {\bibfnamefont {M.}~\bibnamefont
  {Boyle}}, \bibinfo {author} {\bibfnamefont {D.}~\bibnamefont {Iozzo}},\ and\
  \bibinfo {author} {\bibfnamefont {L.~C.}\ \bibnamefont {Stein}},\ }\href@noop
  {} {\bibinfo {title} {{scri} (moble/scri): v1.2}},\ \bibinfo {howpublished}
  {\url{https://github.com/moble/scri}} (\bibinfo {year} {2020}),\ \bibinfo
  {note} {version v1.2; Accessed: 2025-12-20}\BibitemShut {NoStop}%
\bibitem [{\citenamefont {Talbot}\ \emph {et~al.}(2018)\citenamefont {Talbot},
  \citenamefont {Thrane}, \citenamefont {Lasky},\ and\ \citenamefont
  {Lin}}]{PhysRevD.98.064031}%
  \BibitemOpen
  \bibfield  {author} {\bibinfo {author} {\bibfnamefont {C.}~\bibnamefont
  {Talbot}}, \bibinfo {author} {\bibfnamefont {E.}~\bibnamefont {Thrane}},
  \bibinfo {author} {\bibfnamefont {P.~D.}\ \bibnamefont {Lasky}},\ and\
  \bibinfo {author} {\bibfnamefont {F.}~\bibnamefont {Lin}},\ }\href
  {https://doi.org/10.1103/PhysRevD.98.064031} {\bibfield  {journal} {\bibinfo
  {journal} {Phys. Rev. D}\ }\textbf {\bibinfo {volume} {98}},\ \bibinfo
  {pages} {064031} (\bibinfo {year} {2018})}\BibitemShut {NoStop}%
\bibitem [{\citenamefont {Giesler}\ \emph {et~al.}(2019)\citenamefont
  {Giesler}, \citenamefont {Isi}, \citenamefont {Scheel},\ and\ \citenamefont
  {Teukolsky}}]{fit1}%
  \BibitemOpen
  \bibfield  {author} {\bibinfo {author} {\bibfnamefont {M.}~\bibnamefont
  {Giesler}}, \bibinfo {author} {\bibfnamefont {M.}~\bibnamefont {Isi}},
  \bibinfo {author} {\bibfnamefont {M.~A.}\ \bibnamefont {Scheel}},\ and\
  \bibinfo {author} {\bibfnamefont {S.~A.}\ \bibnamefont {Teukolsky}},\ }\href
  {https://doi.org/10.1103/PhysRevX.9.041060} {\bibfield  {journal} {\bibinfo
  {journal} {Phys. Rev. X}\ }\textbf {\bibinfo {volume} {9}},\ \bibinfo {pages}
  {041060} (\bibinfo {year} {2019})}\BibitemShut {NoStop}%
\bibitem [{\citenamefont {Giesler}\ \emph {et~al.}(2025)\citenamefont
  {Giesler}, \citenamefont {Ma}, \citenamefont {Mitman}, \citenamefont
  {Oshita}, \citenamefont {Teukolsky}, \citenamefont {Boyle}, \citenamefont
  {Deppe}, \citenamefont {Kidder}, \citenamefont {Moxon}, \citenamefont
  {Nelli}, \citenamefont {Pfeiffer}, \citenamefont {Scheel}, \citenamefont
  {Throwe},\ and\ \citenamefont {Vu}}]{fit2}%
  \BibitemOpen
  \bibfield  {author} {\bibinfo {author} {\bibfnamefont {M.}~\bibnamefont
  {Giesler}}, \bibinfo {author} {\bibfnamefont {S.}~\bibnamefont {Ma}},
  \bibinfo {author} {\bibfnamefont {K.}~\bibnamefont {Mitman}}, \bibinfo
  {author} {\bibfnamefont {N.}~\bibnamefont {Oshita}}, \bibinfo {author}
  {\bibfnamefont {S.~A.}\ \bibnamefont {Teukolsky}}, \bibinfo {author}
  {\bibfnamefont {M.}~\bibnamefont {Boyle}}, \bibinfo {author} {\bibfnamefont
  {N.}~\bibnamefont {Deppe}}, \bibinfo {author} {\bibfnamefont {L.~E.}\
  \bibnamefont {Kidder}}, \bibinfo {author} {\bibfnamefont {J.}~\bibnamefont
  {Moxon}}, \bibinfo {author} {\bibfnamefont {K.~C.}\ \bibnamefont {Nelli}},
  \bibinfo {author} {\bibfnamefont {H.~P.}\ \bibnamefont {Pfeiffer}}, \bibinfo
  {author} {\bibfnamefont {M.~A.}\ \bibnamefont {Scheel}}, \bibinfo {author}
  {\bibfnamefont {W.}~\bibnamefont {Throwe}},\ and\ \bibinfo {author}
  {\bibfnamefont {N.~L.}\ \bibnamefont {Vu}},\ }\href
  {https://doi.org/10.1103/PhysRevD.111.084041} {\bibfield  {journal} {\bibinfo
   {journal} {Phys. Rev. D}\ }\textbf {\bibinfo {volume} {111}},\ \bibinfo
  {pages} {084041} (\bibinfo {year} {2025})}\BibitemShut {NoStop}%
\bibitem [{\citenamefont {Stein}(2019)}]{steinqnm}%
  \BibitemOpen
  \bibfield  {author} {\bibinfo {author} {\bibfnamefont {L.~C.}\ \bibnamefont
  {Stein}},\ }\href {https://doi.org/10.21105/joss.01683} {\bibfield  {journal}
  {\bibinfo  {journal} {J. Open Source Softw.}\ }\textbf {\bibinfo {volume}
  {4}},\ \bibinfo {pages} {1683} (\bibinfo {year} {2019})},\ \Eprint
  {https://arxiv.org/abs/1908.10377} {arXiv:1908.10377 [gr-qc]} \BibitemShut
  {NoStop}%
\bibitem [{\citenamefont {Oshita}(2021)}]{oshita1}%
  \BibitemOpen
  \bibfield  {author} {\bibinfo {author} {\bibfnamefont {N.}~\bibnamefont
  {Oshita}},\ }\href {https://doi.org/10.1103/PhysRevD.104.124032} {\bibfield
  {journal} {\bibinfo  {journal} {Phys. Rev. D}\ }\textbf {\bibinfo {volume}
  {104}},\ \bibinfo {pages} {124032} (\bibinfo {year} {2021})}\BibitemShut
  {NoStop}%
\bibitem [{\citenamefont {Yi}\ \emph {et~al.}(2024)\citenamefont {Yi},
  \citenamefont {Kuntz}, \citenamefont {Barausse}, \citenamefont {Berti},
  \citenamefont {Cheung}, \citenamefont {Kritos},\ and\ \citenamefont
  {Maselli}}]{Yi:2024}%
  \BibitemOpen
  \bibfield  {author} {\bibinfo {author} {\bibfnamefont {S.}~\bibnamefont
  {Yi}}, \bibinfo {author} {\bibfnamefont {A.}~\bibnamefont {Kuntz}}, \bibinfo
  {author} {\bibfnamefont {E.}~\bibnamefont {Barausse}}, \bibinfo {author}
  {\bibfnamefont {E.}~\bibnamefont {Berti}}, \bibinfo {author} {\bibfnamefont
  {M.~H.-Y.}\ \bibnamefont {Cheung}}, \bibinfo {author} {\bibfnamefont
  {K.}~\bibnamefont {Kritos}},\ and\ \bibinfo {author} {\bibfnamefont
  {A.}~\bibnamefont {Maselli}},\ }\href
  {https://doi.org/10.1103/PhysRevD.109.124029} {\bibfield  {journal} {\bibinfo
   {journal} {Phys. Rev. D}\ }\textbf {\bibinfo {volume} {109}},\ \bibinfo
  {pages} {124029} (\bibinfo {year} {2024})}\BibitemShut {NoStop}%
\bibitem [{\citenamefont {Lagos}\ \emph {et~al.}(2025)\citenamefont {Lagos},
  \citenamefont {Andrade}, \citenamefont {Rafecas-Ventosa},\ and\ \citenamefont
  {Hui}}]{Lagos:2024}%
  \BibitemOpen
  \bibfield  {author} {\bibinfo {author} {\bibfnamefont {M.}~\bibnamefont
  {Lagos}}, \bibinfo {author} {\bibfnamefont {T.}~\bibnamefont {Andrade}},
  \bibinfo {author} {\bibfnamefont {J.}~\bibnamefont {Rafecas-Ventosa}},\ and\
  \bibinfo {author} {\bibfnamefont {L.}~\bibnamefont {Hui}},\ }\href
  {https://doi.org/10.1103/PhysRevD.111.024018} {\bibfield  {journal} {\bibinfo
   {journal} {Phys. Rev. D}\ }\textbf {\bibinfo {volume} {111}},\ \bibinfo
  {pages} {024018} (\bibinfo {year} {2025})},\ \Eprint
  {https://arxiv.org/abs/2411.02264} {arXiv:2411.02264 [gr-qc]} \BibitemShut
  {NoStop}%
\bibitem [{\citenamefont {Shi}\ \emph {et~al.}(2024)\citenamefont {Shi},
  \citenamefont {Zhang},\ and\ \citenamefont {Mei}}]{Shi:2024}%
  \BibitemOpen
  \bibfield  {author} {\bibinfo {author} {\bibfnamefont {C.}~\bibnamefont
  {Shi}}, \bibinfo {author} {\bibfnamefont {Q.}~\bibnamefont {Zhang}},\ and\
  \bibinfo {author} {\bibfnamefont {J.}~\bibnamefont {Mei}},\ }\href
  {https://doi.org/10.1103/PhysRevD.110.124007} {\bibfield  {journal} {\bibinfo
   {journal} {Phys. Rev. D}\ }\textbf {\bibinfo {volume} {110}},\ \bibinfo
  {pages} {124007} (\bibinfo {year} {2024})},\ \Eprint
  {https://arxiv.org/abs/2407.13110} {arXiv:2407.13110 [gr-qc]} \BibitemShut
  {NoStop}%
\end{thebibliography}

\end{document}